\documentclass[%
 aip,
 rsi,
 amsmath,amssymb,
 reprint,%
]{revtex4-2}

\usepackage{hyperref}
\usepackage{graphicx}
\usepackage{dcolumn}
\usepackage{siunitx} 
\newcommand{\rom}[1]{\uppercase\expandafter{\romannumeral #1\relax}}
\usepackage[utf8]{inputenc}
\usepackage[T1]{fontenc}
\usepackage{mathptmx}

\begin{document}

\preprint{AIP/123-QED}

\title{An all-digital associated particle imaging system for the 3D determination of isotopic distributions}

% apart from first and last author, list all authors alphabetically
\author{Mauricio Ayllon Unzueta}
\email{mayllon@berkeley.edu}
\altaffiliation[Also at ]{Nuclear Engineering Department, University of California, Berkeley.}
\altaffiliation[Now at ]{Solar system exploration division, NASA Goddard Space Flight Center, Greenbelt, Maryland.}
\author{Bernhard Ludewigt}
\author{Brian Mak}
\author{Tanay Tak}
\author{Arun Persaud}
 \homepage{http://ibt.lbl.gov}
 \email{apersaud@lbl.gov}
\affiliation{%
Acceleration Technology \& Applied Physics\\
Lawrence Berkeley National Laboratory\\
1 Cyclotron Road, CA 94720, USA
}%

\date{\today}

\begin{abstract}
 Associated particle imaging (API) is a non-destructive nuclear technique for the 3D determination of isotopic distributions. By detecting the alpha particle associated with the emitted neutron in the deuterium-tritium fusion reaction with a position- and time-resolving detector, the direction of the \SI{14.1}{\MeV} neutron and its time of emission can be determined. Employing this method, isotope characteristic gamma rays emitted in inelastic neutron scattering events can be correlated with the neutron interaction location. An API system consisting of a sealed-type neutron generator, gamma detectors, and a position-sensitive alpha detector was designed, constructed, and characterized. The system was tested with common soil elements and shown to be sensitive to $^{12}$C, $^{16}$O, $^{28}$Si, $^{27}$Al, and $^{56}$Fe. New aspects of our approach are the use of a yttrium-aluminum-perovskite (YAP) scintillator, using a sapphire window instead of a fiber-optic faceplate for light transport to the photomultiplier, and the all-digital data acquisition system. We present a description of the system with simulations and experimental results that show a position resolution on the alpha detector of \SI{1}{\milli \meter}, a depth resolution using a LaBr$_3$ detector of \SI{6.2}{\centi \meter}, and an angular resolution of \SI{4.5}{\degree}. Additionally, we present single-element gamma response measurements for the elements mentioned above together with a comparison to Monte Carlo simulations (MCNP6).
\end{abstract}

\maketitle

\section{\label{sec:intro} Introduction}

The origins of associated particle imaging (API) can be traced back to the 1950s and 1960s when the associated particle method (APM) was introduced with the main objective of reducing the induced gamma background \cite{APM1958, CRC} by recording gamma rays in coincidence with the associated particle in a DD or DT fusion reaction, i.e., $^3$He or $^4$He, respectively. In an early proof-of-principle paper, \citet{APIBeyerle} obtained 2D projections for different materials (water and table salt), hence showing some of the most important capabilities of this technique.
Since then, API has found many uses in different areas of research and industry, including the detection of illicit drugs,\cite{FONTANA2017279} explosives,\cite{CARASCO2008397} special nuclear material (SNM),\cite{HAUSLADEN2007, Wellington2015-ac} diamond search,\cite{ALEAKHIN20159} and space exploration.\cite{litvak2019} Fast scintillators, fast electronics, and high-resolution position-sensitive photomultiplier tubes have recently enabled both sub-nanosecond time resolution and sub-millimeter position resolution. This combination allows for the possibility of imaging objects with centimeter scale resolution.

In an API system, neutrons are produced by accelerating a charged particle beam consisting of deuterium (D) and tritium (T) ions onto a titanium target where they accumulate and undergo DT fusion reactions within a small surface area of a few millimeters in diameter. The nuclear reaction shown in Eq.~(\ref{eq:reactionDT}) results in a two-particle decay, and hence the neutron and the alpha particle are emitted back-to-back in the center-of-mass reference frame with fixed energies.
\begin{equation}\label{eq:reactionDT}
D + T \rightarrow \alpha\ (\SI{3.5}{\MeV}) + n\ (\SI{14.1}{\MeV}) \qquad    99.96 \%
\end{equation}

\begin{figure}[htbp]
	\includegraphics[trim=0cm 3cm 3cm 0cm, width=\linewidth, clip=true]{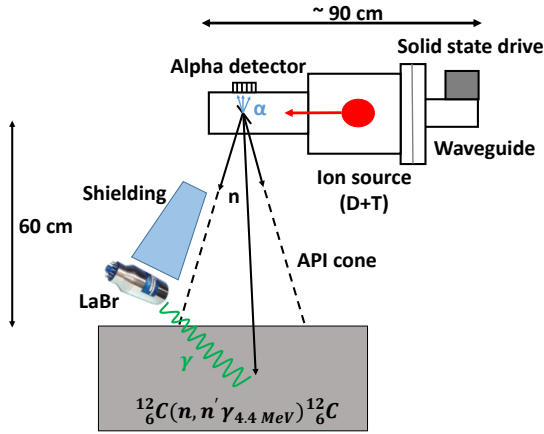}
	\caption{Schematic illustration of the API technique. \SI{14.1}{\MeV} neutrons are produced in a sealed-type neutron generator equipped with a position-sensitive alpha detector. Neutron-induced inelastic gamma rays (e.g.\ from $^{12}$C) are generated in the sample and detected with a LaBr$_3$.}
	\label{fig:setup}
\end{figure}

The detection of the alpha particle in a position-sensitive detector provides information about the direction of the neutron and the time the neutron was emitted. The high-energy neutron can exit the vacuum chamber and excite a nucleus in the assayed sample by inelastic scattering. The de-excitation, which occurs on a picosecond timescale for many isotopes of interest, is accompanied by the emission of one or more gamma rays with an energy characteristic of such an isotope. These gamma rays can be detected with a fast scintillator, and their times and energies are recorded. Based on the time difference between the alpha and gamma detection together with the calculated associated neutron direction, the position of the inelastic scattering reaction can be calculated. By measuring many of these events, 3D elemental density profiles of the object of interest can be reconstructed. This process is exemplified in Figure~\ref{fig:setup}, which shows a particular case where the inelastic scattering in a $^{12}$C nucleus leads to the production of a prompt gamma ray.

The ultimate goal of the API system we are developing is to measure carbon distributions in soil. Non-destructive carbon-in-soil measurement methods are important for understanding and quantifying soil-based carbon sequestration techniques on a large scale. Since soil is the largest storage pool of terrestrial carbon,\cite{Sanderman2017-yz} soil-based carbon sequestration can play an important role in reducing the atmospheric carbon concentration. Monitoring carbon in soil also supports improvements in soil health and crop yield.\cite{Huang2018-md} The standard procedure to measure carbon concentration in soils today consists of taking core (point) samples and analyzing them in a chemistry lab using loss on ignition (LOI) techniques.\cite{LOI} This destructive procedure takes from a few days to weeks before the results are obtained, and results from a few sampling points are then often used to extrapolate the carbon distribution for large areas up to entire fields. The development of this API system is part of an Advanced Research Projects Agency-Energy (ARPA-E) \cite{ROOTS} funded effort to develop better sensors for soil and imaging of roots. This instrument, when fully developed into a commercial system, is expected to be able to non-destructively analyze soil volumes (\SI{50x50x30}{\centi\meter}) \textit{in-situ} within minutes depending on carbon concentration, allowing for better estimates on a field scale. Additionally, its capabilities allow for centimeter resolution in all three dimensions in order to account for carbon heterogeneity. Non-API neutron-based methods for soil characterization have been reported before,\cite{BNLsoil, Wielopolski2008-ff} including instruments that use time-tagged-neutron signals\cite{Kavetskiy2019-rg} and pulsed neutron generators. \cite{Kavetskiy2019-lt} However, non-API/non-time-tagged systems cannot distinguish between different depths and are therefore mostly sensitive to the carbon concentration in the top \SI{8}{\centi\meter}.\cite{Yakubova2016-tk}

In this paper, we focus on the technical aspects of the API system we are developing. Specifically, we show an all-digital data acquisition (DAQ) system in combination with a new approach for alpha detection. For the alpha detector we chose Yttrium Aluminum Perovskite (YAP) as the scintillator material in combination with a sapphire window for light transport to the position-sensitive photomultiplier (PSPMT) instead of the often utilized ZnO and fiber-optic faceplates. In the conclusion section, we briefly mention future improvements and the envisioned path forward for our application of carbon-in-soil measurements. Data shown in this paper was acquired at a lower neutron output than it is possible with the current generator due to limitations in our alpha position readout electronics. In section~\ref{sec:conclusions}, we discuss different methods to achieve operations at higher neutron rates.

\section{\label{sec:description}System overview}

The API system consists of a sealed-type neutron generator, an alpha particle detector, and a gamma-ray detector, i.e., lanthanum bromide (LaBr$_3$). These components are arranged as schematically shown in Figure~\ref{fig:setup}. Our current design allows us to operate a second gamma-ray detector (sodium iodide, NaI) at the same time to achieve higher count rates and shorter measurement times.

The neutron generator is a compact, sealed-type API tube (DT108API, Adelphi Technology~\cite{Adelphi}) with a microwave ion source driven by a solid-state power supply (Sairem GMS200WSM56) which couples to the plasma chamber via a square waveguide with a three-stub tuner. The titanium target, where neutrons are produced, is located inside the vacuum chamber centered underneath the YAP scintillator (manufactured by Crytur \cite{Crytur}) and a 6-inch sapphire window (MPF Products, Inc.). This arrangement allows  the scintillation photons produced by the alpha particles striking the YAP crystal to be transported to the outside of the vacuum chamber where a PSPMT (Hamamatsu H13700-03 \cite{Hamamatsu}) is used to detect the light signals. For the measurements reported in this paper, a LaBr$_3$ gamma detector (Saint-Gobain, 3-inch, B390S) is positioned outside the tagged-neutron cone at the same height as the interrogated samples, as shown in Figure~\ref{fig:experimental-setup}(a). The detector is shielded from the direct neutron flux by 12 inches of high-density polyethylene placed between the generator and the detector. 

The detector itself is surrounded by lead bricks on four sides. The surface of the detector facing the interrogated samples is covered with a thin lead shield (\SI{2} {\milli \meter}). The surface facing away from the samples is not shielded, allowing for signal and high voltage cable connections. Finally, the data acquisition is performed with a 16-channel fully digital system (XIA, PIXIE-16 Rev. F, \SI{500}{\mega\Hz} ADC with \SI{100}{\mega\Hz} FPGAs\cite{XIA}). Signals from the gamma detector and from the penultimate dynode of the alpha detector photomultiplier tube are recorded in coincidence. They are time-stamped using a digital Constant Fraction Discrimination (CFD) algorithm, and pulse heights are recorded in list mode. The coincidence signal also triggers the recording of four corner channels of the position readout of the alpha detector. Therefore, a total of six signals from both the alpha detector (timing + four corners) and gamma detector are preamplified and fed into the PIXIE-16 for digitization and processing.   

The system development described in this paper aims at a proof-of-principle demonstration. A commercial system for carbon-in-soil measurements would require more gamma detectors, the operation of the neutron generator at significantly higher output rates, and an alpha detector readout capable of handling rates on the order of $\SI{1e7}{\alpha\per\second}$. 

\section{\label{sec:alpha}Position-sensitive alpha detector}

The alpha detector allows tagging neutrons emitted into a cone, known as the API cone, and to determine the direction of the neutron and its time of flight. The design goal for the intended application of a final position resolution of \SI{5}{\centi \meter} at a distance of \SI{60}{\centi \meter} from the source led to the following requirements for the alpha detector position and time resolutions. Given the design value of the neutron producing beam spot diameter of $\sim$\SI{2}{\milli\meter} and a distance of about \SI{6}{\centi \meter} between the neutron production site and the alpha detector, a \SI{~1}{\milli \meter} position resolution on the alpha detector is required based on geometry considerations. The alpha-gamma time resolution needs to be \SI{1}{\nano \second} because a \SI{14.1}{\MeV} neutron travels approximately \SI{5}{\centi \meter \per \nano \second}. Furthermore, since the YAP scintillator is located only \SI{6.7}{\centi \meter} from the neutron production site, it experiences a high alpha rate and the system needs to be able to handle these rates. For our geometry, the expected alpha rate is $\SI{1e7}{\alpha\per\second}$ when operating at the maximum generator output of \SI{2e8}{n\per\second}. The design details of the alpha detector and its performance characterization are described at length by \citet{Mauricio1}, and some performance measures are only summarized in this paper. The scintillator signal needs to have a short rise time for optimal time resolution, a fast decay time for reduced pileup at the expected high count rates, and a sufficient light output for accurate position determination. The cerium-doped YAP inorganic scintillator was chosen due to the following reasons: it can withstand vacuum bakeout temperatures, it has a fast response (a time constant of \SI{0.38}{\nano \second} for the rise time has been reported\cite{Moszynski1998-hh}), a short decay time of \SI{27}{\ns}, a high light yield  of approximately \SI{5000}{photons \per \MeV} for alpha particles (\SIrange[range-units = single]{17000}{20000} {photons\per \MeV} for gamma rays), and it has sufficient energy resolution (20\% for \SI{5}{\MeV} alphas). It is also non-hygroscopic and hence easy to handle. The pre-amplified PSPMT signals have a rise time (20\%-80\% value) of \SI{6}{\nano \second} that is well matched to the \SI{500}{\MHz} digitizer. The YAP scintillator is coupled to the PSPMT via a sapphire window vacuum interface. In our current alpha detector setup operating at reduced neutron count rates, all 256 anodes of the PSPMT are resistively connected to a four-corner readout scheme, as discussed in~\citet{Mauricio1} %citet includes a period from et al.
The finite-element software package COMSOL Multyphysics\cite{Comsol} was used to optimize the geometry, understand the light spread onto the photocathode, estimate the total light transmission, and simulate the position reconstruction based on a four-corner readout scheme. The main parameters chosen for the simulations are shown in Table~\ref{tab:comsol-param}.
\begin{table*}[htb]
	\centering
	\begin{ruledtabular}
	\begin{tabular}{llll} 
		\textbf{Name}    & \textbf{Material} & \textbf{Thickness (mm)} & \textbf{Refractive Index (at \SI{370}{\nano \meter})} \\ \hline
		Photocathode     & Bialkali            & {--}   & {--}  \\  
		PSPMT window     & UV glass            & 1.5    & 1.5354\\  
		Optical grease   & PhenylSiO$_2$CH$_3$ & 0.06   & 1.466 \\ 
		Vacuum interface & Sapphire            & 3.0    & 1.7925\\ 
		Vacuum layer     & Vacuum              & 0.005  & 1.0   \\ 
		YAP(Ce)          & YAlO$_3$            & 1.0    & 1.931 \\ 
		Mirror layer     & Aluminum            & 0.0004 & {--}  \\ 
	\end{tabular}
	\end{ruledtabular}
	\caption{\label{tab:comsol-param} Main parameters used in COMSOL simulations of the alpha detector. Most values for the indices of refraction (n) were obtained from the refractive index database.\cite{refractiveIndex}}
\end{table*}
	
\begin{figure}[htbp]
	\includegraphics[trim=0cm 9cm 0cm 0cm, width=\linewidth]{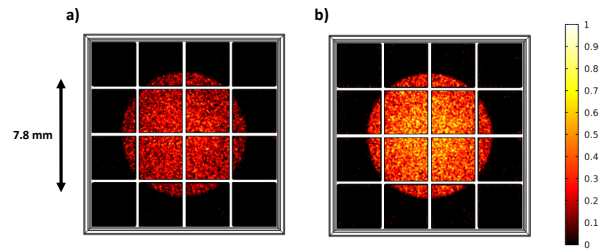}
	\caption{COMSOL simulation of scintillation photon transport through one-sixteenth of the alpha detector system showing (a) the transmitted light with no mirror surface, and (b) the transmitted light with the mirror surface. The transmission increases by a factor of 2 to approximately 6.6\% with no significant change in the diameter of the light projection onto the photocathode. The white lines indicate the detector pixels.}
	\label{fig:comsol-mirror}
\end{figure}

The surface of the YAP facing the incoming flux of alpha particles is coated with a reflective aluminum layer with a thickness of \SI{400}{\nano \meter} to improve photon statistics, which results in an improved position resolution and avoids a significant reduction in the energy of the alpha particle. Figure~\ref{fig:comsol-mirror} shows the COMSOL simulation results of one alpha particle interaction and its resulting light spread onto the photocathode with and without the reflective surface. The simulation shows that on the order of 600 photoelectrons per \SI{3.5}{\MeV} alpha particle will be created in the PSPMT.\cite{Mauricio1}

\begin{figure}[htbp]
	\includegraphics[width =\linewidth]{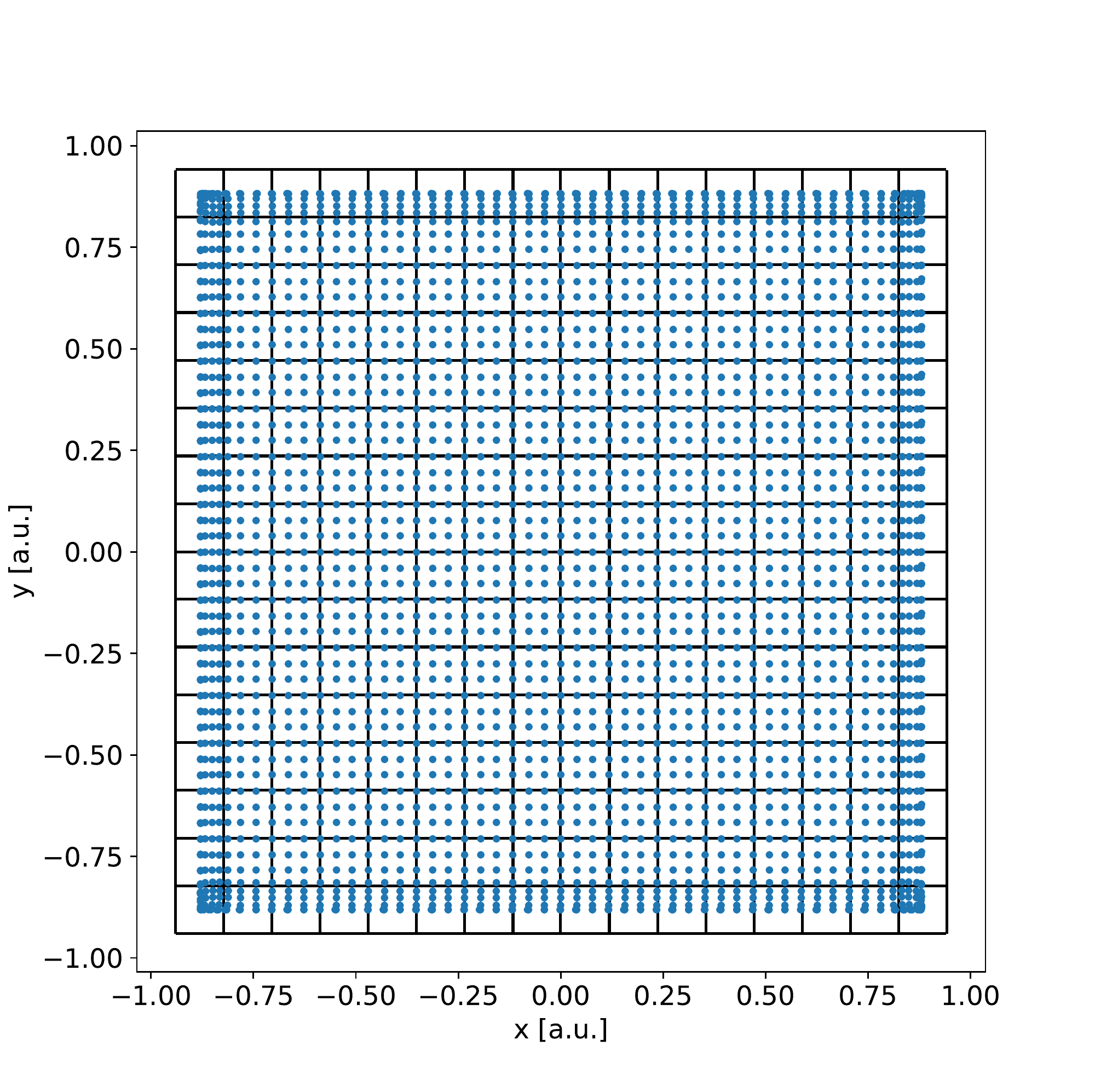}
	\caption{LTspice output showing the resulting calculated position (blue dots) from a distributed DC current source. The source distribution derived from the COMSOL light collection simulations was applied every \SI{0.9}{\milli \meter} and used to simulate the charge collection on neighboring anodes following an alpha particle hit. The reconstruction was done using Eq.~(\ref{eq:xy}). The black-lined squares are the PSPMT pixels. Simulations show a high degree of uniformity except near the edges.}
	\label{fig:spice-4-corners}
\end{figure}

The spread of the scintillation light over several pixels makes it possible to determine the center-of-mass position with a sub-pixel resolution using the following reconstruction algorithm
\begin{equation} \label{eq:xy}
\begin{aligned}
x &= \frac{A+B}{A+B+C+D}\,\,, \\ 
y &= \frac{A+C}{A+B+C+D}\,\,,
\end{aligned}
\end{equation}
where A, B, C, and D are the signal amplitudes read at each corner of the four-corner readout scheme. This algorithm works well for most of the detector area but introduces positional errors near the edges of the detector, which collect only a fraction of these photons. This effect was studied using the simulation software LTspice,\cite{LTSpice} and the results are shown in Figure~\ref{fig:spice-4-corners}. This edge effect can be corrected with more advanced reconstruction techniques taking the partial light collection into account. This will be reported on in a future paper. 

In order to experimentally quantify the position resolution and uniformity of the alpha detector, aluminum masks with different hole patterns were placed in between an $^{241}$Am (\SI{3.9}{\mega \becquerel}) alpha source and the YAP crystal in a separate 6-inch cube vacuum chamber. $^{241}$Am decays primarily by the emission of an alpha particle with an average energy of approximately \SI{5.486}{\MeV}, which is higher than the alpha energy of \SI{3.5}{\MeV} from the DT reaction. Hence, the position reconstruction is expected to be the same, but the higher photon yield of the $^{241}$Am results in lower statistical variations than for the DT case. 

\begin{figure}[htbp]
	\includegraphics[trim=0cm 4cm 0cm 3cm, width =\linewidth]{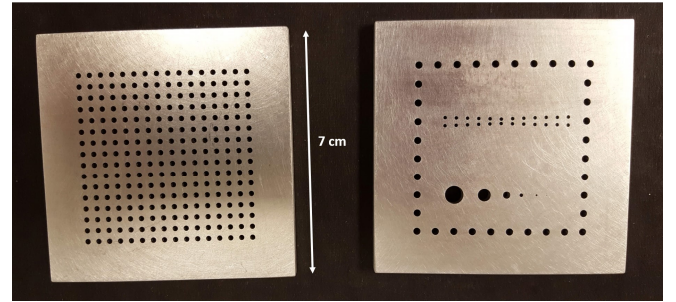}
	\caption{Different aluminum mask designs used for the alpha detector performance tests. The flood-field mask on the left has 256 apertures with a \SI{3}{\mm} center-to-center separation, the same separation as for the PSPMT pixels. The mask on the right was designed to determine the position resolution of the detector.}
	\label{fig:masks-machined}
\end{figure}

\begin{figure}[htbp]
	\includegraphics[trim=0cm 11cm 0cm 2cm, width=\linewidth]{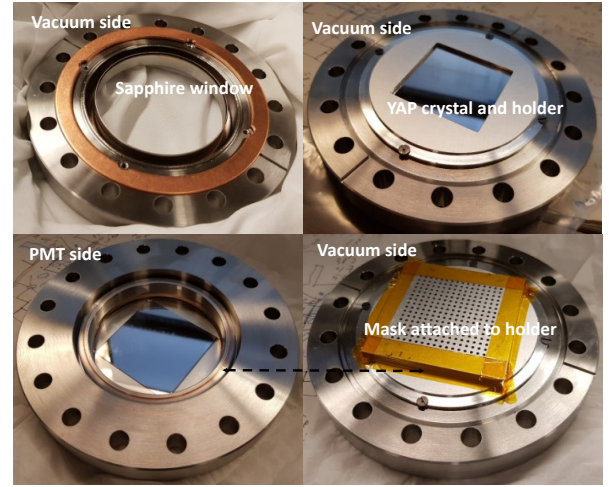}
	\caption{Setup used to test the response of the YAP crystals, the uniformity of the reconstruction algorithm, and the achievable resolution. The sapphire window, YAP, and aluminum holder shown are the same as the ones mounted on the neutron generator.}
	\label{fig:window-mask}
\end{figure}

Figure~\ref{fig:masks-machined} (left) shows a mask with 256 apertures separated by \SI{3}{\mm} center-to-center, which is the same as the separation between individual pixels in the PSPMT.  
These two masks were placed close to the YAP crystal and \SI{10}{\centi \meter} away from a $^{241}$Am source. This system was placed inside a vacuum chamber evacuated to a few mTorr, and data was taken for \SI{200}{\second}. The fast timing signal from the penultimate dynode was used as a trigger for recording the four corner signals. The 6~inch flange with the sapphire window, the YAP holder, and one of the masks are shown in Figure~\ref{fig:window-mask}.

\begin{figure}[htbp]
	\includegraphics[width = \linewidth]{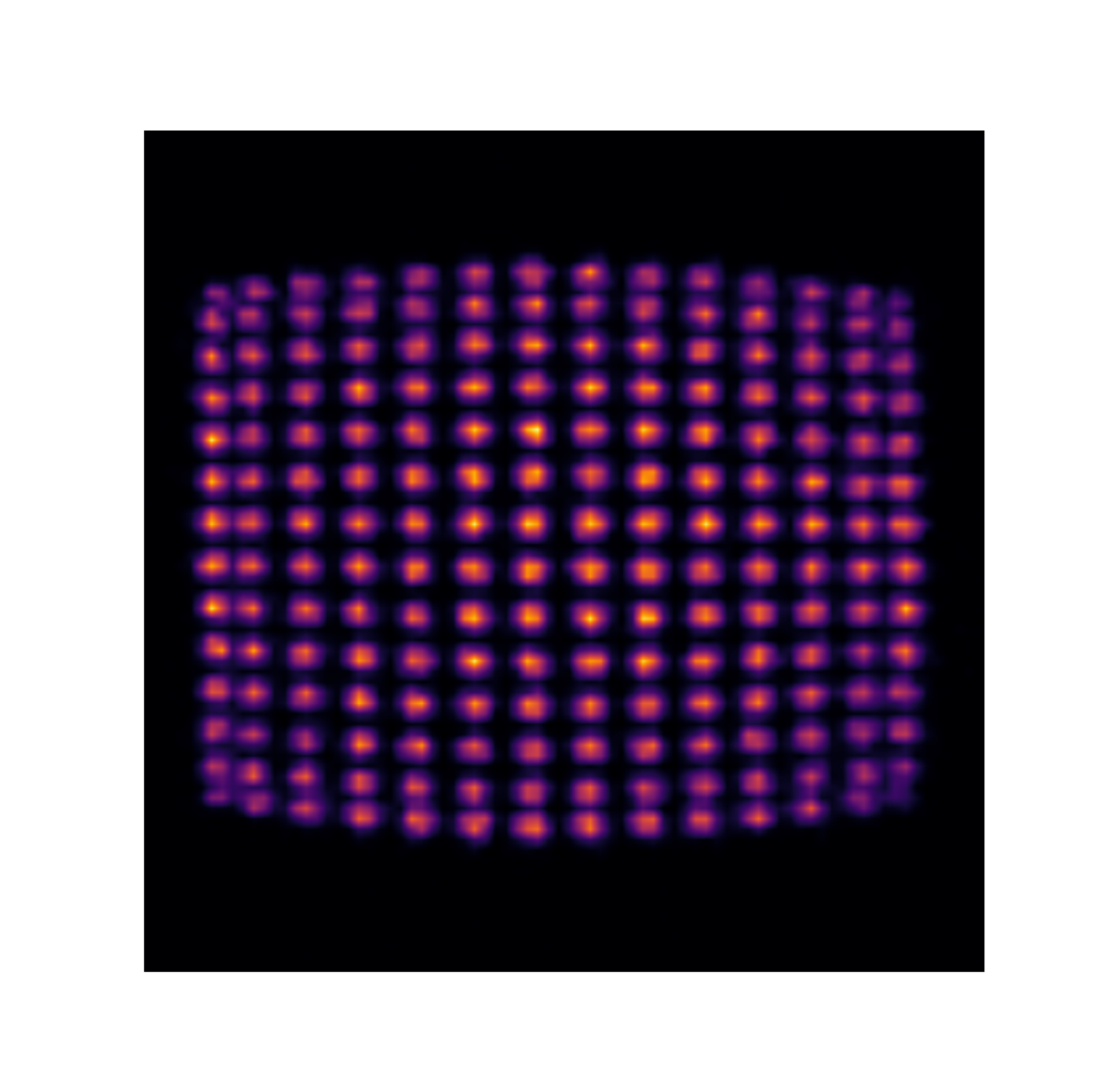}
	\caption{Position reconstruction of the mask configuration shown on the left of Figure~\ref{fig:masks-machined} using Eq.~(\ref{eq:xy}).}
	\label{fig:flood-field-roots}
\end{figure}

The position reconstruction was performed using Eq.~(\ref{eq:xy}). The resulting image corresponding to the 256-hole mask is shown in Figure~\ref{fig:flood-field-roots}. Note the high spatial uniformity except near the edges where light spread is not uniform as predicted in the LTspice simulations above. Additionally, we observe a small non-linear effect (see curvature at the top and bottom) of each reconstructed image that we believe is due to stray capacitance on the readout board.

Additionally, notice how even though a 16$\times$16 mask was used, only a 14$\times$14 array is observed in Figure~\ref{fig:flood-field-roots}, as predicted by COMSOL and LTspice simulations shown in Figure~\ref{fig:spice-4-corners}, which shows this non-linear edge effect.

\begin{figure}[htbp]
	\includegraphics[trim=0cm 9cm 0cm 0cm, width = \linewidth]{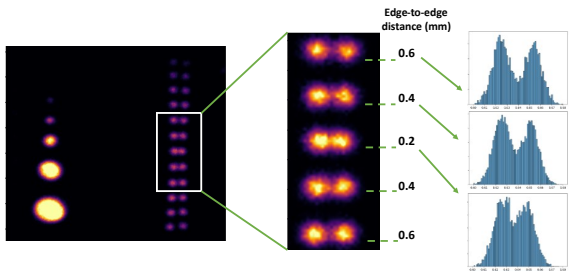}
	\caption{Position resolution test of our detector system showing a series of holes separated by varying distances. Note that the minimum separation between the holes (edge-to-edge) is \SI{0.2}{\mm} and that these can be clearly resolved. A fit using simulated data gives a FWHM of \SI{0.8}{\mm}.}
	\label{fig:mask2_results}
\end{figure}
Figure~\ref{fig:mask2_results} shows the experimental results of using the mask on the right of Figure~\ref{fig:masks-machined} with a hole pattern with decreasing center-to-center distance specially designed to measure the position resolution of the system. Note that even apertures that are \SI{0.2}{\mm} apart (edge-to-edge) can be resolved with our current setup. We then simulated the data by assuming uniform distributed alpha particles within two round areas with the same dimensions and distances as the whole pattern. By adding Gaussian noise to the particles' 2D positions, converting to a 1D histogram, and then fitting to the measured data, we can optimize the full width half maximum (FWHM) of the noise distribution to best fit the measurement. This resulted in a fitted FWHM of \SI{0.8}{\mm} (not shown, but available in Ref.~\onlinecite{data}), which is comparable with results from the work of~\citet{Zhang2012} %citet includes a period from et al.
Scaling the position resolution measured with alphas from a $^{241}$Am source to the lower energy alphas from the DT reaction using the inverse of the square root of the energy (photon statistics) indicates a resolution of \SI{1}{\mm} for the API system.

We therefore achieved the required position resolution as stated in the introduction of this section. However, note that the system's angular resolution also depends on the beam spot size on the neutron target.

\section{\label{sec:daq}PIXIE-16 data acquisition system}

The PIXIE-16 is an all-digital signal processing system that operates at a digitizer speed of \SI{500}{\MHz}. Some of its capabilities regarding energy and time determination in the context of our application are described below. 

\subsection{\label{subsec:energy}Energy determination}

While the PIXIE-16 implements a filter design that allows for pileup correction, for the set of measurements presented in this article, the PIXIE-16 was set to integrate the full energy peak with baseline subtraction. We found that this approach works well for low event rates (alphas and gammas). For reference, using a standard $^{22}$Na radioactive source, we obtained an energy resolution of 3\% at \SI{511}{\kilo \electronvolt} for the LaBr$_3$ detector, which agrees well with the data provided by the manufacturer.

\subsection{\label{subsec:time}Time determination}

The arrival times of the alpha particle and the gamma ray, which are used to calculate the neutron time of flight, need to be measured with excellent resolution for calculating the depth of interaction. As noted previously, achieving the goal of a depth resolution of \SI{5}{\centi \meter} requires a time resolution of approximately \SI{1}{\nano \second}. 
A digital constant fraction discrimination (CFD) algorithm is implemented in the PIXIE-16 for determining the pulse arrival time. The advantage of the CFD method is a much smaller signal time walk obtained by determining the arrival time at a constant fraction of the pulse amplitude. The CFD algorithm implemented in this version of the PIXIE-16 is as follows:
\begin{equation}\label{eq:CFD}
\begin{aligned}
CFD(k) = w & \bigg(\sum_{i=k}^{k+L} a(i) - \sum_{i=k-B}^{k-B+L} \!\!a(i) \bigg) - \\& \bigg(\sum_{i=k-D}^{k-D+L} a(i) - \sum_{i=k-D-B}^{k-D-B+L} \!\!a(i) \bigg),
\end{aligned}
\end{equation} 
This equation has four free parameters, $w$, $B$, $D$, and $L$. The constant fraction $w$ varies between 0 and 1, and the other three are integer values, indicating a number of points in the digitized trace, denoted by $a(i)$. 

\begin{figure}[htbp]
	\includegraphics[width=\linewidth,clip=true]{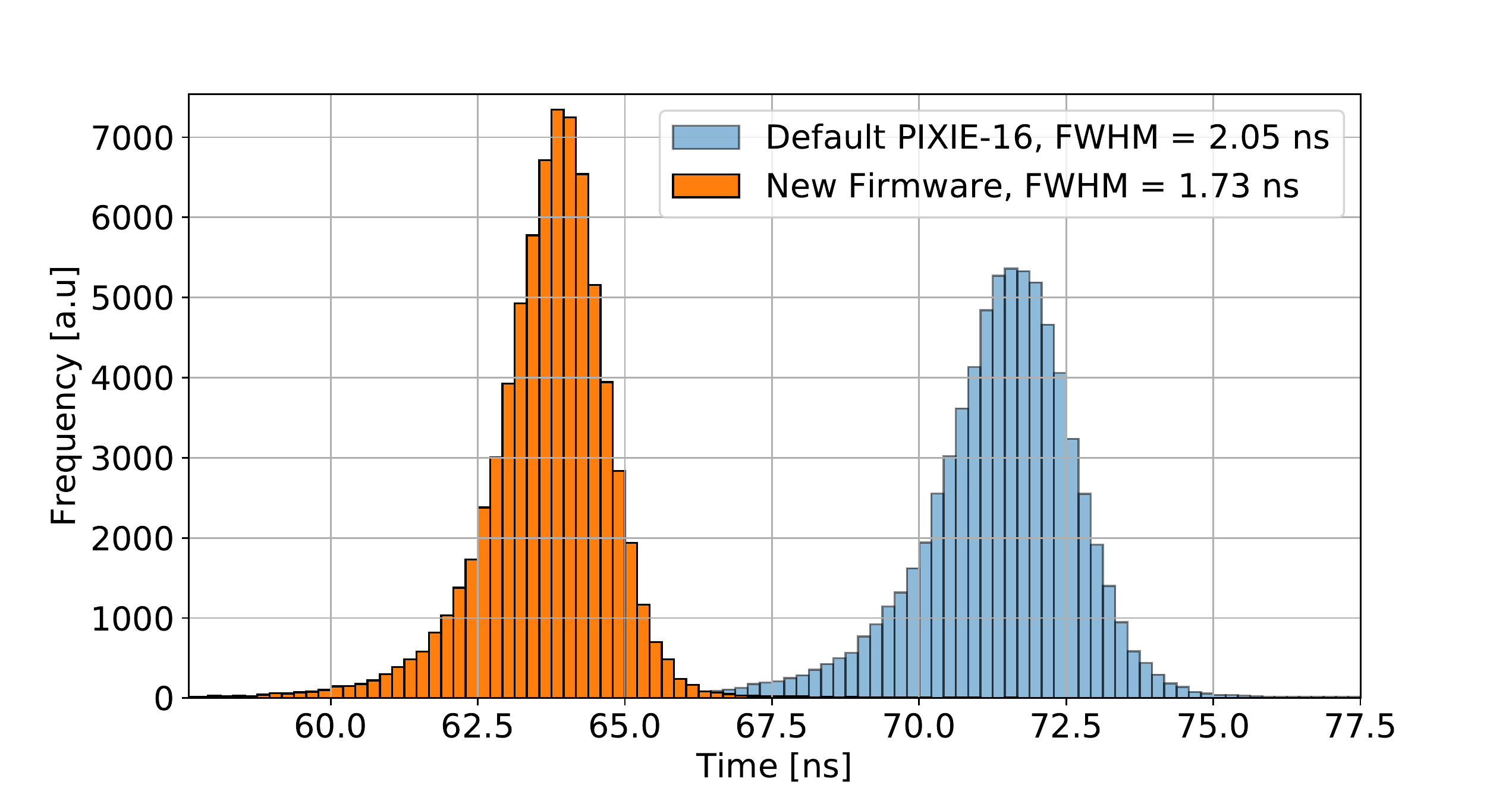}
	\caption{CFD calculation for default PIXIE-16 parameters compared to the case where only $w$ is changed from 1 to 0.3125 for the YAP-LaBr$_3$ combination. Note the time resolution improvement of approximately \SI{300}{\pico\second}. Note: the time resolution of the actual API system is significantly better due to the higher signal amplitudes generated by the alpha particles and higher energy gamma rays.}
	\label{fig:Na22-new-firmware}
\end{figure}

Due to the implementation details of the CFD algorithm in the FPGA of the PIXIE-16, the parameters cannot be arbitrarily chosen. For example, the integration times are constrained to multiples of five because the clock frequency of the FPGAs (\SI{100}{\mega\Hz}) is five times lower than the digitizer frequency (\SI{500}{\mega\Hz}). The parameters were optimized offline for the three different detectors of the API system. We analyzed coincident signal traces collected with a $^{22}$Na source that emits two \SI{511}{\kilo \electronvolt} gamma rays simultaneously. The results indicated that the parameter that makes the most significant change is $w$, the multiplicative factor by which the trace amplitude is reduced by a certain fraction. The time resolution of the system improves as $w$ decreases. However, at the same time, the peak of the CFD trace that is used for triggering decreases in amplitude, and hence, the CFD threshold has to be set closer to the noise level, which sets a lower limit on $w$. We picked a near optimal value of $w=0.3125$ that could easily be implemented in a custom firmware. As can be seen in Figure~\ref{fig:Na22-new-firmware}, a time resolution of \SI{1.73}{\nano \second} was achieved using the $^{22}$Na source, which is an improvement of \SI{300}{\pico\second} compared to the standard firmware with a $w$ value of 1. The actual API system has an overall better time resolution due to higher alpha and gamma energies which provide improved statistics. The results shown below indicate a Z-resolution of \SI{6.2}{\centi\meter}; using the neutron time of flight, one can calculate an upper bound for the time resolution of \SI{1.25}{\nano\second}. We expect the time resolution of the system to be better than this value because the value reported above also includes effects of neutron scattering and reconstruction errors.

The appropriate time windows to be implemented in the DAQ logic depend on the arrival times of the coincident alpha particle and gamma ray signals. The main components are the flight times of the neutron and alpha particles (geometry dependent) and the electron transit times of the respective photomultiplier tubes. Experimental data indicate a total time delay between the alpha (YAP) and gamma (LaBr$_3$) signals of approximately \SI{80}{\nano \second} (of which roughly \SI{10}{\nano\second} are due to the time of flight). The pre-processed and time-stamped data are recorded and used for further analysis. While alpha hits in each pixel are not recorded, the total number of alpha events (ignoring the coincidence setup) is recorded in order to calculate alpha/neutron rates and normalize the neutron output between runs.

\section{\label{sec:api}API system performance}

The overall performance of API systems can be quantified in terms of its position resolution in 3D (or equivalently using angular resolution for X-Y) and its ability to obtain a prompt gamma-ray spectrum from a specific volume in a given time. The measurement time will vary depending on the nature and size of the interrogated sample, the number of gamma detectors employed for the measurement, the rate capability of the alpha detector, and the neutron rate. For the carbon-in-soil application, the measurement time will ultimately be limited by a maximum neutron rate of about \SI{2e8}{n\per \second} where accidental coincidences between gamma and alphas will start dominating the signal.\cite{Mauricio1} In the following sections, we present measurements and simulation results to quantify the angular resolution and depth resolution of the current system. Additionally, we show the measured gamma response to specific samples (pure elements) that highlight the ability of API to select certain sub-volumes of the target area and suppress background gamma rays.

\subsection{\label{subsec:reconstruction} API reconstruction algorithm}
\begin{figure}[htbp]
	\includegraphics[width=\linewidth]{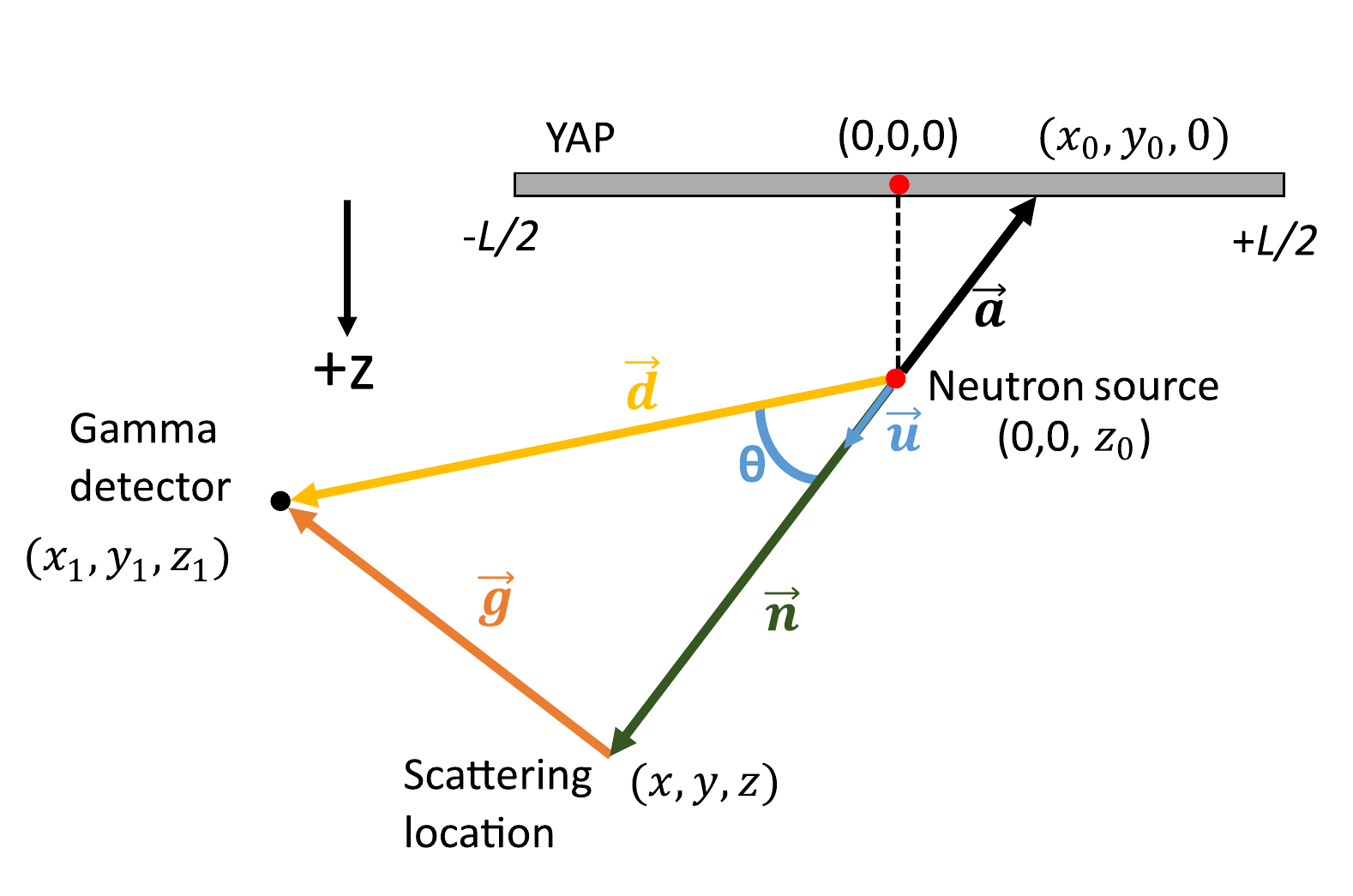}
	\caption{Schematic representation (not to scale) of the API reconstruction technique based on vector algebra.}
	\label{fig:API-reconstruction}
\end{figure}

The pre-processed data from the DAQ are used to reconstruct the 3D location of an inelastic neutron scattering (INS) event followed by the emission of a prompt gamma ray. The ($x_0, y_0$) position on the YAP crystal of the alpha detector is calculated from the four-corner energies, as explained in Section~\ref{sec:alpha}. Figure~\ref{fig:API-reconstruction} illustrates how the scattering location ($x$,$y$,$z$) in the sample is then calculated based on vector algebra.
If we assume a neutron point source located at a known distance from the YAP crystal, we can calculate the direction and velocity vector of the alpha particle, $\vec{a}$. Assuming that the neutron travels in the opposite direction of that alpha particle and together with the measured time difference between the alpha and gamma arrival, we can calculate the position of the scattering location by solving a quadratic equation that takes the time of flight of the gamma and neutron into account. The mathematical details of the algorithm can be found in Ref.~\onlinecite{thesisMauricio}.

% moved the figure up to force it to show up on top of page 7
\begin{figure*}[tb!]
	\includegraphics[width=0.9\linewidth,clip=true]{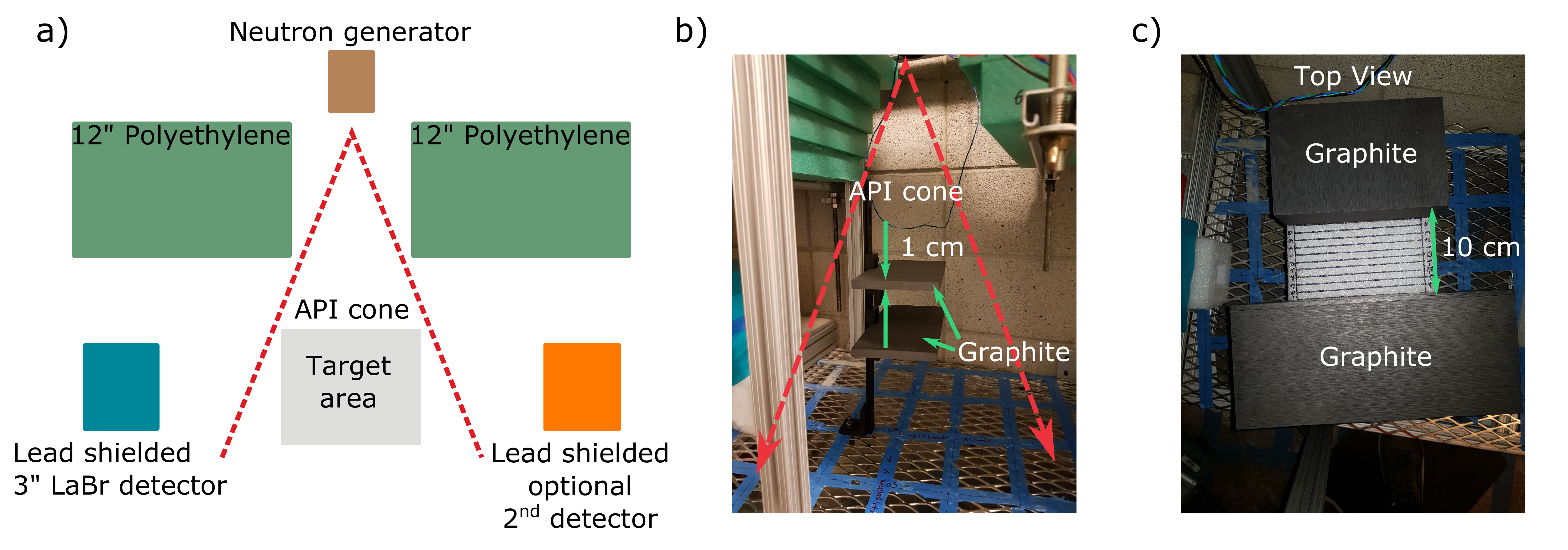}
	\caption{Experimental setup used to characterize the system: (a) schematic of the setup showing the position of the neutron source, the shielding, and the detectors, (b) depth resolution measurement using thin (\SI{1}{\centi \meter}) graphite slabs, and (c) X-Y resolution using thick (\SI{6}{\centi \meter}) graphite bricks.}
	\label{fig:experimental-setup}
\end{figure*} 

The main assumptions in the reconstruction are that the neutrons are monoenergetic (\SI{14.1}{\MeV}) and emitted from a point source, the neutron and associated alpha are emitted exactly \SI{180}{\degree} from each other in the lab system, and the position of the gamma detector is a point in space. Additionally, the alpha, gamma, and neutron are taken to have fixed velocities for subsequent calculations. All of these assumptions are approximations of the actual system, and each of them contributes to the overall error in the position reconstruction. In particular, the fact that we ignore the center-of-mass velocity will add a systematic shift in the reconstruction of several degrees depending on the center-of-mass velocity (i.e., acceleration voltage). However, to a first order approximation, this effect will only affect the reconstruction in the direction of the beam ($x$-axis) and will be a systematic shift, which does not affect the resolution. For a more accurate reconstruction, an average center-of-mass velocity needs to be taken into account and we plan to do so in the future.

In our system, the main contributions to the angular uncertainty of the emitted neutrons are twofold: the size of the ion beam spot on the neutron generating target of approximately \SI{2}{\milli \meter} at \SI{100}{\kilo \volt} and the position resolution of the alpha detector of $<$ \SI{1}{\milli \meter}. The angular resolution in this case is mostly determined by the beam spot size and is expected to be approximately \SI{3}{\degree} based on the nominal beam spot size and measured alpha detector resolution. Other contributions to this uncertainty include multiple scattering of the alpha particles in the target material and the dependence of the angle between the alpha and the neutron on the energy of the incoming deuteron (or triton) that decreases as it penetrates deeper into the target material before interacting with a triton (or deuteron). Both of these contributions are smaller than the geometric uncertainty (at \SI{50}{\kilo \volt} they are significantly smaller), and for a \SI{3}{\degree} angular uncertainty, one expects an X-Y resolution at \SI{60}{\centi \meter} distance of about \SI{3}{\centi \meter}. The depth resolution is dominated by the timing uncertainty, given that a \SI{14.1}{\MeV} neutron travels approximately \SI{5}{\centi \meter\per\nano\second}.
In the following discussion, we report on experimentally measured spatial resolution of the system and compare the results with simulations where we added Gaussian noise in order to represent the system resolution.

\subsection{\label{subsec:resolution} Position resolution}

In order to test the system resolution in three dimensions, we performed a series of experiments using graphite bricks (99\% $^{12}$C) where the event of interest is the detection of the single INS gamma ray of \SI{4.439}{\MeV}. The graphite samples were arranged in two different configurations: (1) two thin (\SI[product-units = power]{1.27x12.5x25}{\centi \meter}) slabs stacked in Z with varying distances between them in order to measure the depth resolution, as shown in Figure~\ref{fig:experimental-setup}(b) and (2) two thick (\SI[product-units = power]{20.4x6.6x12.9}{\centi \meter}), (\SI[product-units = power]{33x5.9x14.3}{\centi \meter}) graphite bricks placed parallel to each other (X-Y plane) and their distance varied along the X dimension, as shown in Figure~\ref{fig:experimental-setup}(c). We used thicker bricks to measure the X-Y resolution simply because they yield a higher gamma count rate, and hence, the measurement time decreases. The neutron generator was operated at \SI{50}{\kilo \volt}, producing a neutron output of $\approx \SI{5e6}{n \per \second}$. Note that we operated at this reduced voltage because we are limited by the alpha rate that our detector can handle. In section~\ref{sec:conclusions}, we discuss different options to handle higher neutron and alpha rates up to \SI{2e8}{n\per\second}, which correspond to the maximum acceleration voltage of our system (\SI{100}{\kilo \volt}).

\begin{figure*}[h!tb]
	\includegraphics[width=0.9\linewidth, clip=true]{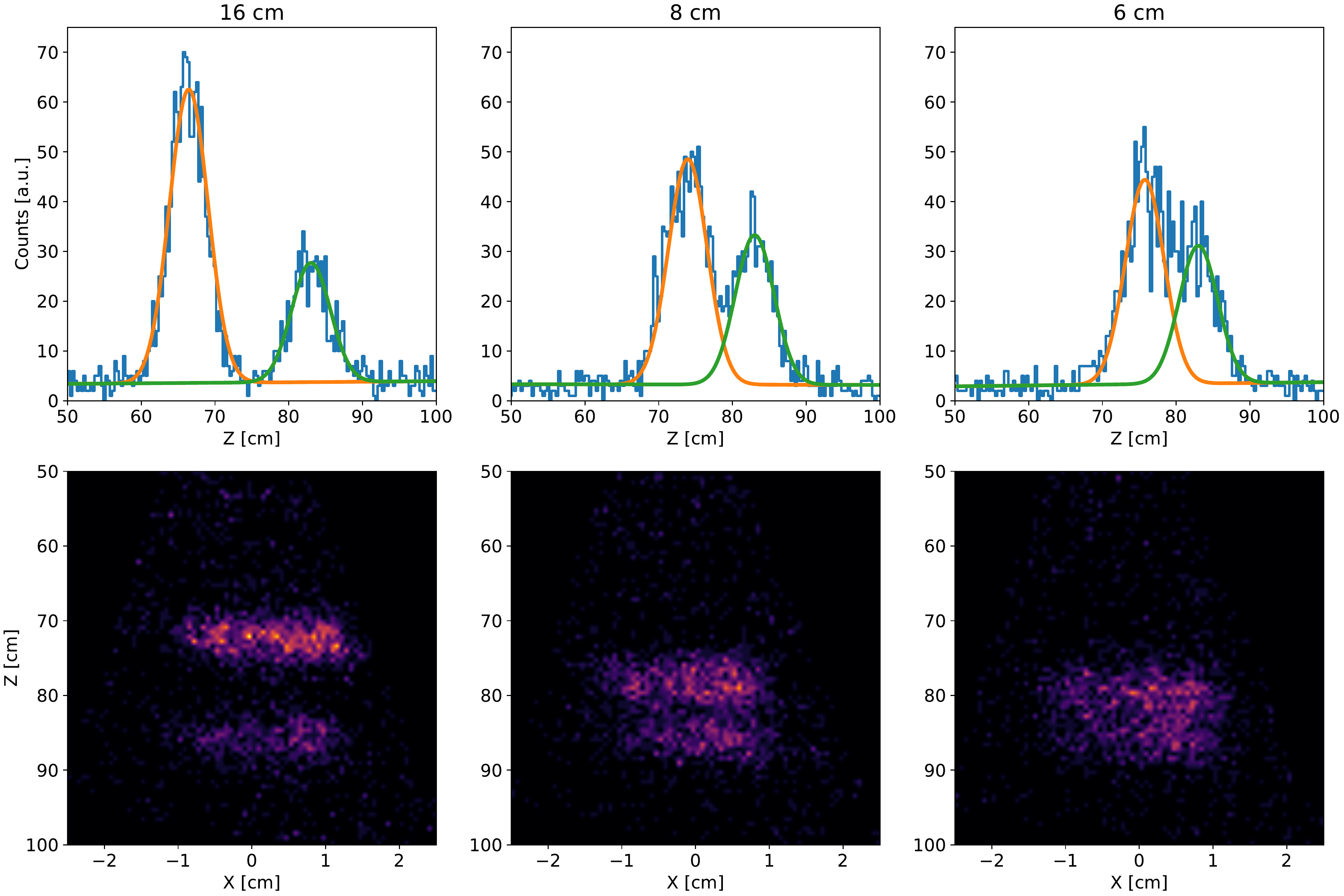}
	\caption{LaBr$_3$ experimental results for the depth resolution measurements with two thin graphite slabs at varying distances. The bottom plots show X-Z intensity maps of the 3D locations of the reconstructed events, and the top plots show the integrated counts vs Z with Gaussian curves used to fit the two peaks. Note that at \SI{6}{\centi \meter} separation, the two slabs are still distinguishable. The intensity maps also show the top slab being exposed to a higher neutron flux as expected.}
	\label{fig:z-res-labr}
\end{figure*} 
Figure~\ref{fig:z-res-labr} shows the experimental results for three different cases where the top graphite slab was brought closer to the bottom one starting with a \SI{10}{\centi \meter} separation between them. The irradiation time for each case scenario was \SI{3600}{\second}. The data was processed by selecting events in the range of the carbon peak at a gamma energy of \SI{4.439}{\MeV} and its corresponding single escape peak (e.g., \SIrange{3.8}{4.5}{\MeV}). Additionally, we only selected counts in the projection of the graphite area in the X-Y plane. We define the intrinsic depth resolution of the system as the FWHM of the Gaussian fit of the measured Z-profile of a single graphite sample (the contribution of the thickness of the graphite slab can be ignored). A depth resolution of \SI{6.2 +- 0.1}{\centi \meter} was determined for the LaBr$_3$-YAP detector combination from the Gaussian fits shown in Figure~\ref{fig:z-res-labr}. This also agrees well with the measurements shown in the same figure of the two graphite bricks at different distances, where we can distinguish the individual bricks down to a distance of approximately \SI{6}{\centi\meter}.

The system X-Y resolution was measured similarly where two thick graphite bricks were placed side by side separated initially by \SI{10}{\centi\meter}. The  bricks  were  then  brought  together in \SI{1}{\centi\meter} steps. They were oriented such that their main axis was parallel to the ion beam and to the main axis of the gamma-ray detector. The irradiation time for each case scenario was \SI{3000}{\second}. 
\begin{figure*}[htb!]
	\includegraphics[width=0.9\linewidth, clip=true]{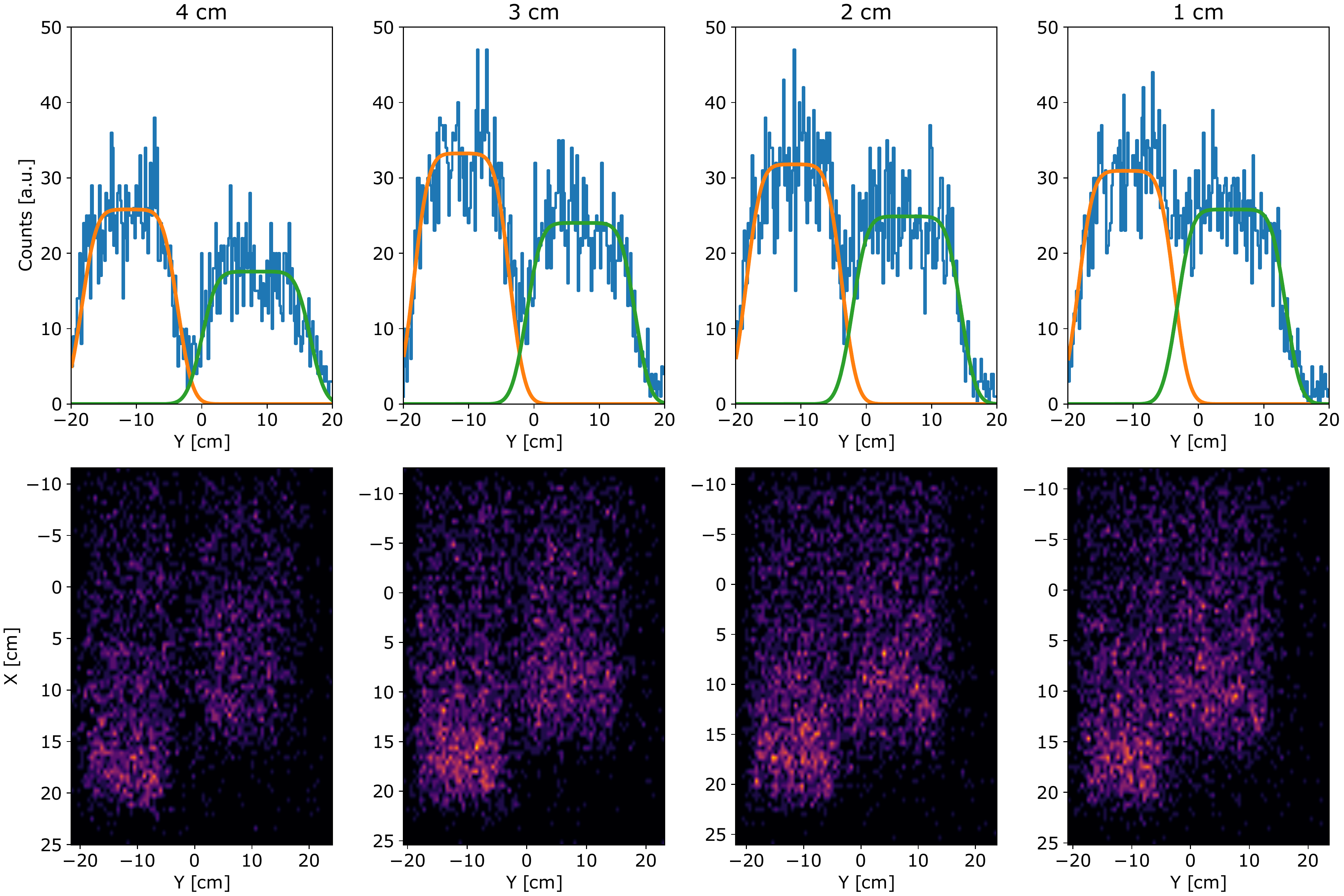}
	\caption{Experimental results for the X-Y resolution of the system measured with graphite bricks and a LaBr$_3$ gamma-ray detector. The bottom plots show X-Y intensity maps of the 3D locations of the reconstructed events, and the top plots show the integrated counts vs X with the sum of 200 Gaussian curves used to fit the two peaks.}
	\label{fig:xy-res-labr-experiment}
\end{figure*}
Because the bricks used for this experiment were thicker, their widths cannot be ignored in this case and we used a sum over multiple Gaussians (all with the same FWHM) to obtain a fit. The FWHM obtained from these measurement is \SI{4.2 +- 0.2}{\centi \meter} at \SI{60}{\centi\meter}, corresponding to an angular resolution of \SI{4}{\degree}.

In Figure~\ref{fig:xy-res-labr-experiment}, the graphite bricks seem to be distinguishable down to a distance of \SI{2}{\centi\meter}. However, as we will show in the next section, this is due to the thickness of the graphite bricks being substantially larger than the X-Y resolution. To confirm this, we simulated the experiment using MCNP6, a Monte-Carlo radiation transport simulation code.\cite{MCNP}

\subsection{\label{subsec:simulations} Simulations}

We recreated the same scenarios as in the X-Y experiments in MCNP6 to produce a simulated dataset with the corresponding geometry and placement of the neutron source, alpha detector, carbon bricks, and gamma detectors. The relevant difference between the measured and simulated dataset was that in the simulation, we forced the main assumptions of our reconstruction algorithm to be true, namely, that the neutrons and associated alphas were generated at a single point in space and that the alpha detector had perfect position resolution. Furthermore, the simulation does not produce any timing errors. Therefore, we consider the reconstructed X-Y distribution of the simulated dataset to have no error. We then added random Gaussian errors to the timing distribution and to the constructed position on the alpha detectors. 

We varied the standard deviation of these distributions to fit the measured data. We also fitted the target-to-scintillator distance because the target position had to be adjusted during the construction of the neutron generator. Furthermore, our reconstruction algorithm has a free parameter for each gamma detector that accounts for timing offsets of the digitizers that are introduced when rebooting the PIXIE-16. We also allowed for a X-Y offset between the simulated and laboratory coordinate systems to adjust for any errors in the coordinate system that was established during the experiment.
\begin{figure*}[bht!]
	\includegraphics[width=0.95\linewidth, clip=true]{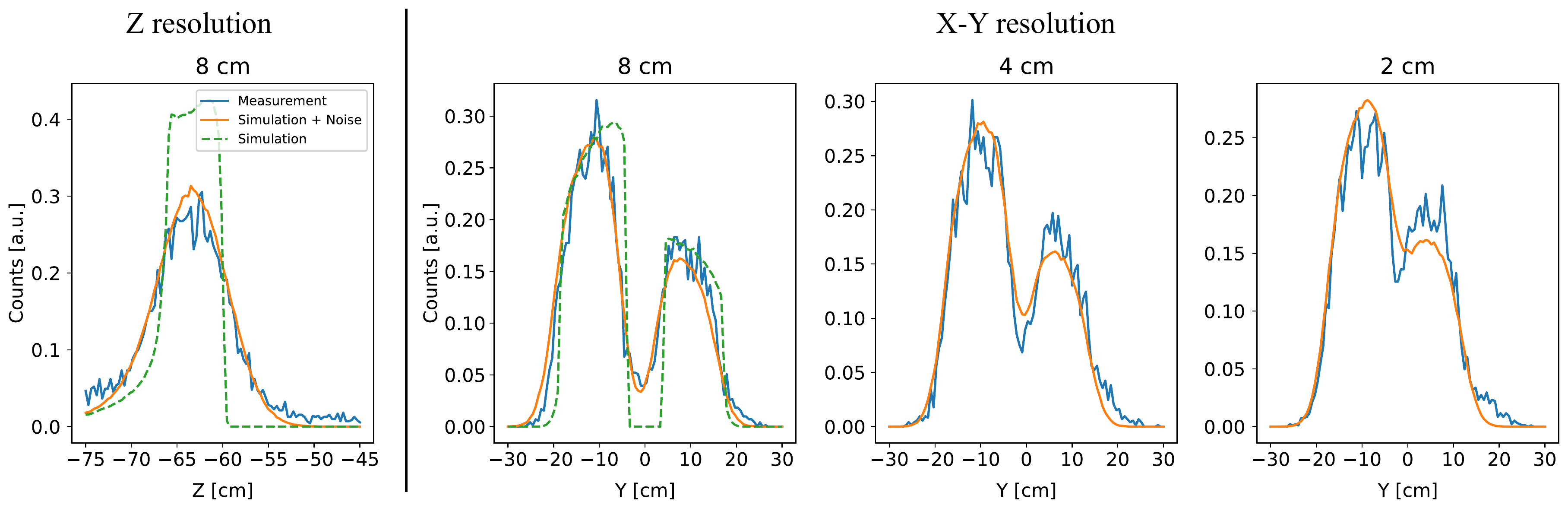}
	\caption{Simulated MCNP6 data shown for different experimental distances between the graphite bricks. The plots for \SI{8}{\centi\meter} distances also show the raw reconstructed simulated distributions assuming no errors. The standard deviation of the added noise was optimized to fit the measured results.}
	\label{fig:xy-res-labr-simulation}
\end{figure*}

The results of the simulation are shown in Figure~\ref{fig:xy-res-labr-simulation} and show good agreement between the measured data and the simulation. The plots for a distance of \SI{8}{\centi\meter} between the graphite bricks show the distribution of the raw simulated data (without added Gaussian noise) and the fitted data (with noise). The \SI{8}{\centi \meter} case also shows the simulated data without additional noise, and one can clearly see the effect of the additional noise rounding off the edges of the larger graphite bricks. For the X-Y case of two bricks, this gives the appearance of a larger distance between the bricks than the actual gap.

The added Gaussian X-Y noise at the location of the alpha detector that resulted in the best fit was \SI{5.3}{\milli\meter} (FWHM), and the estimated distance between the alpha detector and neutron source was \SI{67}{\milli\meter}, which gives an angular resolution of \SI{4.5}{\degree} for our system. The FWHM of the timing noise that fitted the Z-distribution best was \SI{1.25}{\nano\second}. The simulation results for the angular resolution agree within half a degree with our earlier results of fitting the measured Y-distribution directly with Gaussians. The fitted timing noise also agrees well with our earlier estimate from the $^{22}$Na measurement, as well as the measured FWHM of \SI{6.2}{\centi \meter} from the Z-resolution measurement.

\subsection{\label{subsec:gamma} Single-element gamma response}

As mentioned previously, one of the most important characteristics of the API technique is the ability to obtain prompt gamma spectra from a specific volume so that background gammas from surrounding materials or delayed emission (such as from neutron capture) are greatly reduced. To demonstrate this using our instrument, we measured single-element spectra of the main elements in soil, which also served to validate our MCNP6 models. 
\begin{figure}[htb]
	\includegraphics[width=\linewidth, clip=true]{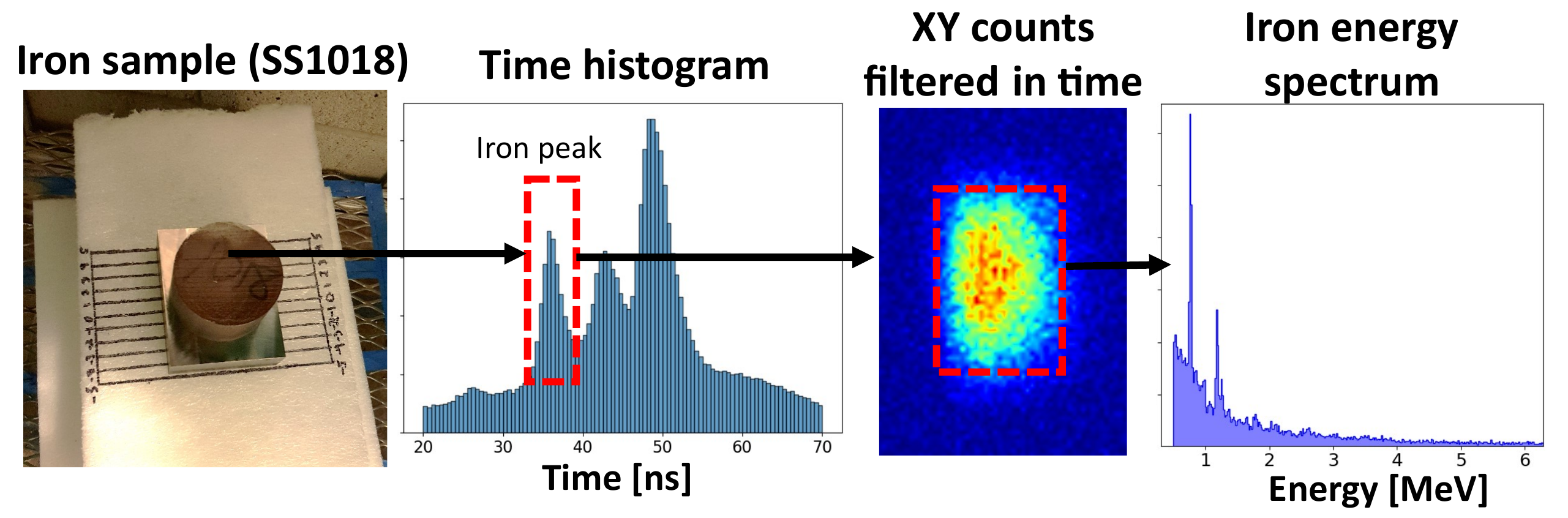}
	\caption{Example of the analysis procedure for obtaining elemental gamma spectra for specific elements. The photograph of the sample is a top view from the perspective of the neutron generator. The LaBr$_3$ detector is located to the left of it.}
	\label{fig:elements-Fe-example}
\end{figure}
Figure~\ref{fig:elements-Fe-example} shows an iron sample (SS1018) of \SI{3.16}{\kg} that was irradiated for \SI{100}{\minute} using a neutron generator voltage of \SI{50}{\kilo \volt}. The sample was placed approximately \SI{60}{\centi \meter} underneath the neutron generator. From the measured data X, Y, and Z coordinates are calculated and the events are restricted to the volume of the sample under test. An energy histogram of all the events inside this volume is then generated. The inelastic gamma spectrum of natural iron as measured is shown on the right of Figure~\ref{fig:elements-Fe-example}.
   
The same procedure was used to obtain element-specific gamma spectra for various elements identified as the most abundant in forest and agricultural soils: carbon, aluminum, oxygen, iron, and silicon.

The measured spectra were generally linear up to the carbon peak (\SI{4.4}{\MeV}) but showed some non-linearity at higher gamma-ray energies. Therefore, we used a linear calibration function up to \SI{5}{\MeV}, and from 5 to \SI{10}{\MeV}, we used a quadratic function, which resulted in a better agreement with expected peak positions. 
The measured elemental energy spectra served to benchmark our MCNP6 simulations. Simulations are generally needed in order to design the instrument, plan future experiments, and understand the sensitivity of the system to different elemental concentrations. They can be used to optimize specific design parameters and can support the analysis of experimental data. Here, we simulated the experiment in two steps: (1) Neutron-induced gamma-ray production in the sample and (2) gamma-ray transport to the detector with its specific response function.
\begin{figure}[htbp]
	\includegraphics[width=\linewidth, clip=true]{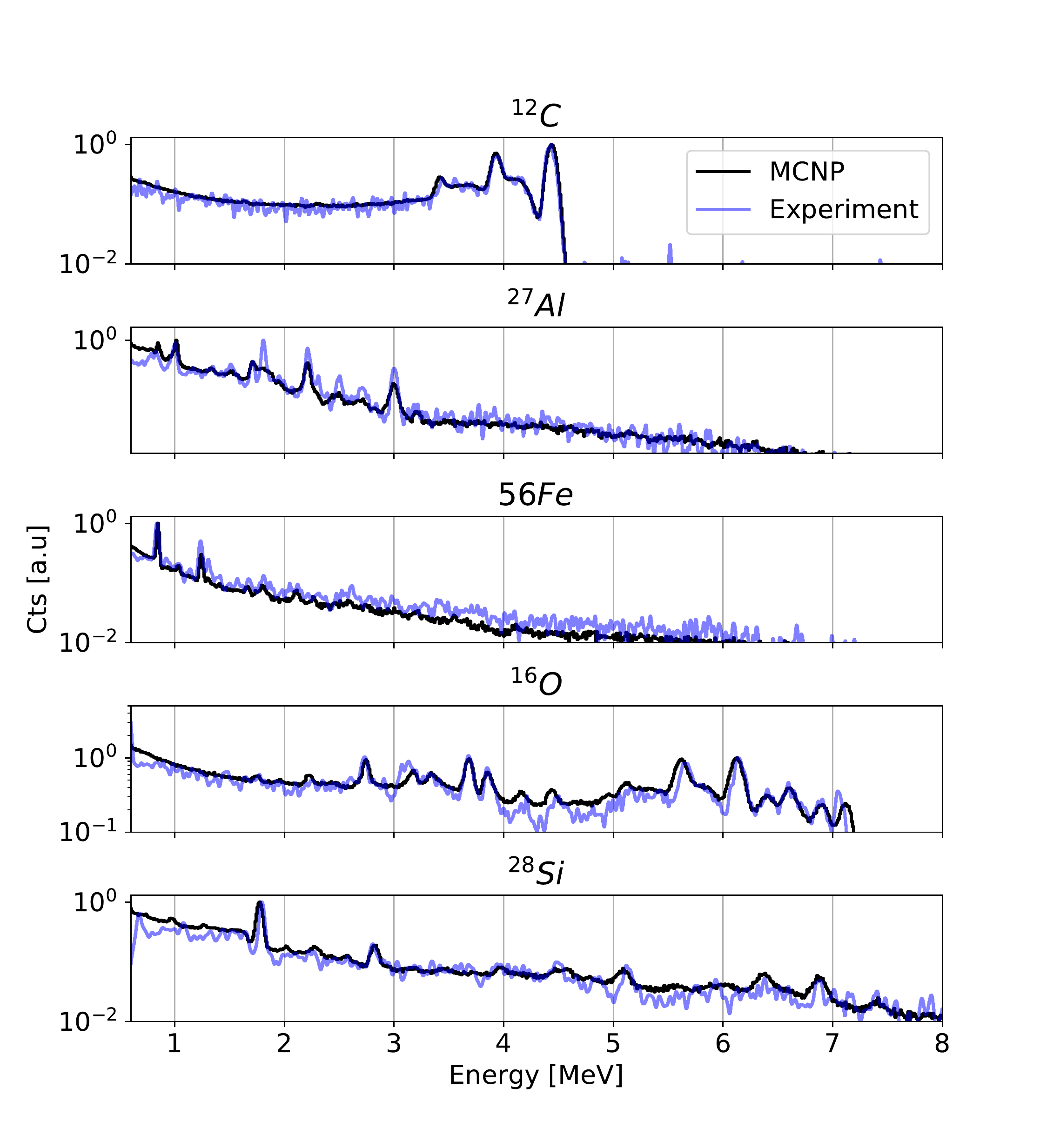}
	\caption{Comparison between measured gamma-ray spectra and MCNP6 simulations for elements relevant to soil composition. Note the overall agreement. However, there are significant discrepancies for $^{27}$Al, in particular. The spectra were normalized to the highest peak.}
	\label{fig:MCNP-exp-all}
\end{figure}
Figure~\ref{fig:MCNP-exp-all} shows the measured spectra with their corresponding simulation. While there is generally good agreement, there are some significant discrepancies in relative intensities such as for the \SI{1.72}{\MeV} gamma-ray line from $^{27}$Al(n,n'$\gamma$)$^{27}$Al originating from the transition \SI{2.73}{\MeV} $\rightarrow$ \SI{1.01}{\MeV}. The nuclear data library ENDF~/B-\rom{6} was used because of its better agreement with experiments compared to most current releases of ENDF.\cite{ENDF} The reason seems to be the attempt to transition from natural compounds to individual isotopes in later versions, which created gaps and overall poorer agreement with experimental data.\cite{Mauborgne} The results presented in Figure~\ref{fig:MCNP-exp-all} show the capability of the system to analyze samples within a small volume and obtain their gamma signature. 
   
\section{\label{sec:conclusions}Results and Conclusions}

A prototype of an all-digital associated particle imaging system for the determination of isotopic 3D distributions in soils has been designed, built, and tested. We found that our approach of coupling a monolithic YAP crystal to a sapphire vacuum window gave similar resolution and higher light yields than previously developed approaches using fiber-optic faceplates, such as in the works of \citeauthor{Cates2013}.\cite{Cates2013}
Experimental tests performed with an $^{241}$Am source gave a \SI{0.8}{\milli \meter} position resolution for the emitted alpha particles of \SI{5.486}{\MeV} using a four-corner readout scheme. Scaling the measured resolution with $1/\sqrt{E_{\alpha}}$ we estimate a position resolution of \SI{1}{\milli \meter} for the detection of \SI{3.5}{\MeV} alphas from DT reactions. 

The API system was characterized in terms of its position resolution and its ability to identify specific isotopes relevant to soil chemistry. We demonstrated a system angular resolution of \SI{4.5}{\degree} and a depth (Z) resolution of \SI{6.2 +- 0.1}{\centi \meter} using graphite samples. 

The alpha detector position resolution of \SI{1}{\milli\meter} and the nominal beam spot size (\SI{2}{\milli\meter}) should allow us to achieve a \SI{3}{\degree} angular resolution. We attribute the measured lower resolution to a larger beam spot on the neutron target. This seems reasonable since we are operating at a considerable lower voltage (\SI{50}{\kilo\volt}) than the design voltage (\SI{100}{\kilo\volt}), resulting in non-optimal ion beam optics and likely a larger beam spot on the neutron producing target.

The depth resolution is mainly dependent on the time resolution, and we achieved noticeable improvements by implementing a custom firmware that optimizes the CFD parameters. Further optimizing the CFD parameters or switching to a different algorithm to measure the arrival time of gamma rays and alphas\cite{Warburton2017-qe} could improve the depth resolution even further.

We obtained gamma-ray spectra from different elemental materials, including $^{12}$C, $^{16}$O, $^{28}$Si, $^{27}$Al, and $^{56}$Fe, and used these data to benchmark Monte Carlo simulations. We found generally good agreement, confirming that the radiation transport code MCNP6 can be used to optimize the system and support the data analysis. The simulation of the X, Y, and Z data also agrees well with the measured data. 

Currently, the maximum neutron rate at which the system can be operated is limited by pileup on the four-corner readout signals of the alpha detector. Data presented in this paper was taken at a neutron rate of approximately \SI{5e6}{n\per\second}. However, the current system already works with minimal position degradation at rates up to \SI{1e7}{n\per\second} (not shown). Reducing the RC time-constants introduced by the resistive network and utilizing the built-in pileup correction of the PIXIE-16 are expected to increase the rate capability of the four-corner readout further. However, additional measurements are needed to understand the extend to which pileup can be rejected without significantly degrading the alpha detector position resolution.

The currently used four-corner readout logic is relatively easy to implement and needs only few digitizer channels but has two major downsides. First, it introduces an RC delay that broadens the PMT signal and leads to increased pileup. Second, every corner experiences the full event rate because all 256 pixels are connected to each corner.

We believe that the path to operate at higher rates of \SI{2e8}{n\per\second} is to digitize every single pixel individually. This will remove the RC delay and will also reduce the event rate per channel. Developments in digitizer and FPGA boards are making the goal of digitizing all 256 channels reachable. An intermediate step could be to digitize single rows and columns, requiring a smaller number of digitizers and allowing for a lower RC constant at the same time. We already designed and built a board for this purpose and plan to test it in the near future. 

In conclusion, we have shown that YAP and a sapphire window present a possible alternative that can achieve high resolution for the alpha detection in an API system, possibly enabling a \SI{3}{\degree} angular resolution. A four-corner readout scheme yields positive results at a low neutron rate. In order to operate at higher neutron rates, digitizing single pixels should be implemented. Operating at lower rate showed that clean elemental spectra can be obtained by reconstructing the position of the scattering center and constraining the gamma spectra to a sub-volume in the target area, thereby drastically reducing background noise. We demonstrated an angular resolution of \SI{4}{\degree} and attribute this lower than expected resolution to a larger beam spot size when operating at lower extraction voltage. We think that the angular resolution will improve when operating at higher voltages.

\section{Outlook}
The results presented indicate that the developed API system is already very close to the design goal of measuring carbon distribution in soil with a \SI{5x5x5}{\centi\meter} voxel resolution over a \SI{50x50x30}{\centi\meter} sample volume. However, higher rate capabilities need to be implemented and the timing resolution needs to be slightly improved. Work in progress includes switching to an improved setup,\cite{thesisMauricio} allowing operation at higher neutron rates, better detector shielding, and a portable setup that can be used in outdoor settings. Finally, we also need to implement models to reconstruct soil elemental densities from the raw data taking neutron and gamma attenuation and detector efficiencies into account. Currently, ongoing experiments indicate that our system has a linear response to the carbon concentration in soil samples for the relevant range of \SIrange{0}{10}{\percent} carbon concentration and we will present these results in a future publication.
All experiments so far indicate that our goal of measuring carbon distribution in soil is achievable. However, field tests of the system and benchmarks against known soil samples to verify its applicability are needed.

\section*{Author's contributions}
All authors contributed to writing and editing the paper and to the experimental work. Mauricio Ayllon Unzueta did most of the measurements discussed in this paper and ran all MCNP6 simulations. Brian Mak worked on estimating the angular resolution using simulated data. Tanay Tak worked mostly on the SPICE simulations and the X-Y reconstruction in the alpha detector. Mauricio Ayllon Unzueta, Bernhard Ludewigt, and Arun Persaud worked on the instrument design, experimental setup, data acquisition system, and data analysis. Arun Persaud also worked on the hardware control and is the principal investigator of the project.

\begin{acknowledgments}
The authors would like to thank Takeshi Katayanagi for his technical support. Furthermore, we would like to thank Wolfgang Henning and Hui Tan at XIA for their help.
The information, data, or work presented herein was funded by the Advanced Research Projects Agency-Energy (ARPA-E), U.S. Department of Energy, under Contract No. DEAC02-05CH11231. 
\end{acknowledgments}

\section*{Data Availability}
The data, analysis scripts, and simulation scripts are openly available on Zenodo at \url{https://doi.org/10.5281/zenodo.4737896},
reference number \onlinecite{data}.

\bibliography{references}

%aipnum4-2.bst 2019-01-14 (MD) hand-edited version of apsrev4-1.bst
\providecommand{\noopsort}[1]{}\providecommand{\singleletter}[1]{#1}%
\begin{thebibliography}{35}%
\makeatletter
\providecommand \@ifxundefined [1]{%
 \@ifx{#1\undefined}
}%
\providecommand \@ifnum [1]{%
 \ifnum #1\expandafter \@firstoftwo
 \else \expandafter \@secondoftwo
 \fi
}%
\providecommand \@ifx [1]{%
 \ifx #1\expandafter \@firstoftwo
 \else \expandafter \@secondoftwo
 \fi
}%
\providecommand \natexlab [1]{#1}%
\providecommand \enquote  [1]{``#1''}%
\providecommand \bibnamefont  [1]{#1}%
\providecommand \bibfnamefont [1]{#1}%
\providecommand \citenamefont [1]{#1}%
\providecommand \href@noop [0]{\@secondoftwo}%
\providecommand \href [0]{\begingroup \@sanitize@url \@href}%
\providecommand \@href[1]{\@@startlink{#1}\@@href}%
\providecommand \@@href[1]{\endgroup#1\@@endlink}%
\providecommand \@sanitize@url [0]{\catcode `\\12\catcode `\$12\catcode
  `\&12\catcode `\#12\catcode `\^12\catcode `\_12\catcode `\%12\relax}%
\providecommand \@@startlink[1]{}%
\providecommand \@@endlink[0]{}%
\providecommand \url  [0]{\begingroup\@sanitize@url \@url }%
\providecommand \@url [1]{\endgroup\@href {#1}{\urlprefix }}%
\providecommand \urlprefix  [0]{URL }%
\providecommand \Eprint [0]{\href }%
\providecommand \doibase [0]{https://doi.org/}%
\providecommand \selectlanguage [0]{\@gobble}%
\providecommand \bibinfo  [0]{\@secondoftwo}%
\providecommand \bibfield  [0]{\@secondoftwo}%
\providecommand \translation [1]{[#1]}%
\providecommand \BibitemOpen [0]{}%
\providecommand \bibitemStop [0]{}%
\providecommand \bibitemNoStop [0]{.\EOS\space}%
\providecommand \EOS [0]{\spacefactor3000\relax}%
\providecommand \BibitemShut  [1]{\csname bibitem#1\endcsname}%
\let\auto@bib@innerbib\@empty
%</preamble>
\bibitem [{\citenamefont {Okhuysen}\ \emph {et~al.}(1958)\citenamefont
  {Okhuysen}, \citenamefont {Bennett}, \citenamefont {Ashe},\ and\
  \citenamefont {Millett}}]{APM1958}%
  \BibitemOpen
  \bibfield  {author} {\bibinfo {author} {\bibfnamefont {P.~L.}\ \bibnamefont
  {Okhuysen}}, \bibinfo {author} {\bibfnamefont {E.~W.}\ \bibnamefont
  {Bennett}}, \bibinfo {author} {\bibfnamefont {J.~B.}\ \bibnamefont {Ashe}},\
  and\ \bibinfo {author} {\bibfnamefont {W.~E.}\ \bibnamefont {Millett}},\
  }\bibfield  {title} {\enquote {\bibinfo {title} {Detection of fast neutrons
  by the associated particle method},}\ }\href
  {https://doi.org/10.1063/1.1716073} {\bibfield  {journal} {\bibinfo
  {journal} {Review of Scientific Instruments}\ }\textbf {\bibinfo {volume}
  {29}},\ \bibinfo {pages} {982--985} (\bibinfo {year} {1958})},\ \Eprint
  {https://arxiv.org/abs/https://doi.org/10.1063/1.1716073}
  {https://doi.org/10.1063/1.1716073} \BibitemShut {NoStop}%
\bibitem [{\citenamefont {Csikai}(1987)}]{CRC}%
  \BibitemOpen
  \bibfield  {author} {\bibinfo {author} {\bibfnamefont {J.}~\bibnamefont
  {Csikai}},\ }\href@noop {} {\emph {\bibinfo {title} {CRC Handbook of Fast
  Neutron Generators}}}\ (\bibinfo  {publisher} {CRC, Boca Raton, FL},\
  \bibinfo {year} {1987})\BibitemShut {NoStop}%
\bibitem [{\citenamefont {{Beyerle}}\ \emph {et~al.}(1991)\citenamefont
  {{Beyerle}}, \citenamefont {{Durkee}}, \citenamefont {{Headley}},
  \citenamefont {{Hurley}},\ and\ \citenamefont {{Tunnell}}}]{APIBeyerle}%
  \BibitemOpen
  \bibfield  {author} {\bibinfo {author} {\bibfnamefont {A.}~\bibnamefont
  {{Beyerle}}}, \bibinfo {author} {\bibfnamefont {R.}~\bibnamefont {{Durkee}}},
  \bibinfo {author} {\bibfnamefont {G.}~\bibnamefont {{Headley}}}, \bibinfo
  {author} {\bibfnamefont {J.~P.}\ \bibnamefont {{Hurley}}},\ and\ \bibinfo
  {author} {\bibfnamefont {L.}~\bibnamefont {{Tunnell}}},\ }\bibfield  {title}
  {\enquote {\bibinfo {title} {Associated particle imaging},}\ }in\ \href
  {https://doi.org/10.1109/NSSMIC.1991.259135} {\emph {\bibinfo {booktitle}
  {Conference Record of the 1991 IEEE Nuclear Science Symposium and Medical
  Imaging Conference}}}\ (\bibinfo {year} {1991})\ pp.\ \bibinfo {pages}
  {1298--1304 vol.2}\BibitemShut {NoStop}%
\bibitem [{\citenamefont {Fontana}\ \emph {et~al.}(2017)\citenamefont
  {Fontana}, \citenamefont {Carnera}, \citenamefont {Lunardon}, \citenamefont
  {Pino}, \citenamefont {Sada}, \citenamefont {Soramel}, \citenamefont
  {Stevanato}, \citenamefont {Nebbia}, \citenamefont {Carasco}, \citenamefont
  {Perot}, \citenamefont {Sardet}, \citenamefont {Sannie}, \citenamefont
  {Iovene}, \citenamefont {Tintori}, \citenamefont {Grodzicki}, \citenamefont
  {Moszyński}, \citenamefont {Sibczyński}, \citenamefont {Swiderski},\ and\
  \citenamefont {Moretto}}]{FONTANA2017279}%
  \BibitemOpen
  \bibfield  {author} {\bibinfo {author} {\bibfnamefont {C.~L.}\ \bibnamefont
  {Fontana}}, \bibinfo {author} {\bibfnamefont {A.}~\bibnamefont {Carnera}},
  \bibinfo {author} {\bibfnamefont {M.}~\bibnamefont {Lunardon}}, \bibinfo
  {author} {\bibfnamefont {F.}~\bibnamefont {Pino}}, \bibinfo {author}
  {\bibfnamefont {C.}~\bibnamefont {Sada}}, \bibinfo {author} {\bibfnamefont
  {F.}~\bibnamefont {Soramel}}, \bibinfo {author} {\bibfnamefont
  {L.}~\bibnamefont {Stevanato}}, \bibinfo {author} {\bibfnamefont
  {G.}~\bibnamefont {Nebbia}}, \bibinfo {author} {\bibfnamefont
  {C.}~\bibnamefont {Carasco}}, \bibinfo {author} {\bibfnamefont
  {B.}~\bibnamefont {Perot}}, \bibinfo {author} {\bibfnamefont
  {A.}~\bibnamefont {Sardet}}, \bibinfo {author} {\bibfnamefont
  {G.}~\bibnamefont {Sannie}}, \bibinfo {author} {\bibfnamefont
  {A.}~\bibnamefont {Iovene}}, \bibinfo {author} {\bibfnamefont
  {C.}~\bibnamefont {Tintori}}, \bibinfo {author} {\bibfnamefont
  {K.}~\bibnamefont {Grodzicki}}, \bibinfo {author} {\bibfnamefont
  {M.}~\bibnamefont {Moszyński}}, \bibinfo {author} {\bibfnamefont
  {P.}~\bibnamefont {Sibczyński}}, \bibinfo {author} {\bibfnamefont
  {L.}~\bibnamefont {Swiderski}},\ and\ \bibinfo {author} {\bibfnamefont
  {S.}~\bibnamefont {Moretto}},\ }\bibfield  {title} {\enquote {\bibinfo
  {title} {Detection system of the first rapidly relocatable tagged neutron
  inspection system (rrtnis), developed in the framework of the european h2020
  c-bord project},}\ }\href
  {https://doi.org/https://doi.org/10.1016/j.phpro.2017.09.010} {\bibfield
  {journal} {\bibinfo  {journal} {Physics Procedia}\ }\textbf {\bibinfo
  {volume} {90}},\ \bibinfo {pages} {279 -- 284} (\bibinfo {year} {2017})},\
  \bibinfo {note} {conference on the Application of Accelerators in Research
  and Industry, CAARI 2016, 30 October – 4 November 2016, Ft. Worth, TX,
  USA}\BibitemShut {NoStop}%
\bibitem [{\citenamefont {Carasco}\ \emph {et~al.}(2008)\citenamefont
  {Carasco}, \citenamefont {Perot}, \citenamefont {Bernard}, \citenamefont
  {Mariani}, \citenamefont {Szabo}, \citenamefont {Sannie}, \citenamefont
  {Roll}, \citenamefont {Valkovic}, \citenamefont {Sudac}, \citenamefont
  {Viesti}, \citenamefont {Lunardon}, \citenamefont {Bottosso}, \citenamefont
  {Fabris}, \citenamefont {Nebbia}, \citenamefont {Pesente}, \citenamefont
  {Moretto}, \citenamefont {Zenoni}, \citenamefont {Donzella}, \citenamefont
  {Moszynski}, \citenamefont {Gierlik}, \citenamefont {Batsch}, \citenamefont
  {Wolski}, \citenamefont {Klamra}, \citenamefont {Tourneur}, \citenamefont
  {Lhuissier}, \citenamefont {Colonna}, \citenamefont {Tintori}, \citenamefont
  {Peerani}, \citenamefont {Sequeira},\ and\ \citenamefont
  {Salvato}}]{CARASCO2008397}%
  \BibitemOpen
  \bibfield  {author} {\bibinfo {author} {\bibfnamefont {C.}~\bibnamefont
  {Carasco}}, \bibinfo {author} {\bibfnamefont {B.}~\bibnamefont {Perot}},
  \bibinfo {author} {\bibfnamefont {S.}~\bibnamefont {Bernard}}, \bibinfo
  {author} {\bibfnamefont {A.}~\bibnamefont {Mariani}}, \bibinfo {author}
  {\bibfnamefont {J.-L.}\ \bibnamefont {Szabo}}, \bibinfo {author}
  {\bibfnamefont {G.}~\bibnamefont {Sannie}}, \bibinfo {author} {\bibfnamefont
  {T.}~\bibnamefont {Roll}}, \bibinfo {author} {\bibfnamefont {V.}~\bibnamefont
  {Valkovic}}, \bibinfo {author} {\bibfnamefont {D.}~\bibnamefont {Sudac}},
  \bibinfo {author} {\bibfnamefont {G.}~\bibnamefont {Viesti}}, \bibinfo
  {author} {\bibfnamefont {M.}~\bibnamefont {Lunardon}}, \bibinfo {author}
  {\bibfnamefont {C.}~\bibnamefont {Bottosso}}, \bibinfo {author}
  {\bibfnamefont {D.}~\bibnamefont {Fabris}}, \bibinfo {author} {\bibfnamefont
  {G.}~\bibnamefont {Nebbia}}, \bibinfo {author} {\bibfnamefont
  {S.}~\bibnamefont {Pesente}}, \bibinfo {author} {\bibfnamefont
  {S.}~\bibnamefont {Moretto}}, \bibinfo {author} {\bibfnamefont
  {A.}~\bibnamefont {Zenoni}}, \bibinfo {author} {\bibfnamefont
  {A.}~\bibnamefont {Donzella}}, \bibinfo {author} {\bibfnamefont
  {M.}~\bibnamefont {Moszynski}}, \bibinfo {author} {\bibfnamefont
  {M.}~\bibnamefont {Gierlik}}, \bibinfo {author} {\bibfnamefont
  {T.}~\bibnamefont {Batsch}}, \bibinfo {author} {\bibfnamefont
  {D.}~\bibnamefont {Wolski}}, \bibinfo {author} {\bibfnamefont
  {W.}~\bibnamefont {Klamra}}, \bibinfo {author} {\bibfnamefont {P.~L.}\
  \bibnamefont {Tourneur}}, \bibinfo {author} {\bibfnamefont {M.}~\bibnamefont
  {Lhuissier}}, \bibinfo {author} {\bibfnamefont {A.}~\bibnamefont {Colonna}},
  \bibinfo {author} {\bibfnamefont {C.}~\bibnamefont {Tintori}}, \bibinfo
  {author} {\bibfnamefont {P.}~\bibnamefont {Peerani}}, \bibinfo {author}
  {\bibfnamefont {V.}~\bibnamefont {Sequeira}},\ and\ \bibinfo {author}
  {\bibfnamefont {M.}~\bibnamefont {Salvato}},\ }\bibfield  {title} {\enquote
  {\bibinfo {title} {In-field tests of the euritrack tagged neutron inspection
  system},}\ }\href
  {https://doi.org/https://doi.org/10.1016/j.nima.2008.01.097} {\bibfield
  {journal} {\bibinfo  {journal} {Nucl. Instrum. Methods Phys. Res. A}\
  }\textbf {\bibinfo {volume} {588}},\ \bibinfo {pages} {397 -- 405} (\bibinfo
  {year} {2008})}\BibitemShut {NoStop}%
\bibitem [{\citenamefont {Hausladen}\ \emph {et~al.}(2007)\citenamefont
  {Hausladen}, \citenamefont {Bingham}, \citenamefont {Neal}, \citenamefont
  {Mullens},\ and\ \citenamefont {Mihalczo}}]{HAUSLADEN2007}%
  \BibitemOpen
  \bibfield  {author} {\bibinfo {author} {\bibfnamefont {P.}~\bibnamefont
  {Hausladen}}, \bibinfo {author} {\bibfnamefont {P.}~\bibnamefont {Bingham}},
  \bibinfo {author} {\bibfnamefont {J.}~\bibnamefont {Neal}}, \bibinfo {author}
  {\bibfnamefont {J.}~\bibnamefont {Mullens}},\ and\ \bibinfo {author}
  {\bibfnamefont {J.}~\bibnamefont {Mihalczo}},\ }\bibfield  {title} {\enquote
  {\bibinfo {title} {Portable fast-neutron radiography with the nuclear
  materials identification system for fissile material transfers},}\ }\href
  {https://doi.org/https://doi.org/10.1016/j.nimb.2007.04.206} {\bibfield
  {journal} {\bibinfo  {journal} {Nuclear Instruments and Methods in Physics
  Research Section B: Beam Interactions with Materials and Atoms}\ }\textbf
  {\bibinfo {volume} {261}},\ \bibinfo {pages} {387 -- 390} (\bibinfo {year}
  {2007})},\ \bibinfo {note} {the Application of Accelerators in Research and
  Industry}\BibitemShut {NoStop}%
\bibitem [{\citenamefont {Wellington}\ \emph {et~al.}(2015)\citenamefont
  {Wellington}, \citenamefont {Palles}, \citenamefont {Mullens}, \citenamefont
  {Mihalczo}, \citenamefont {Archer}, \citenamefont {Thompson}, \citenamefont
  {Britton}, \citenamefont {Ezell}, \citenamefont {Ericson}, \citenamefont
  {Farquhar}, \citenamefont {Lind},\ and\ \citenamefont
  {Carter}}]{Wellington2015-ac}%
  \BibitemOpen
  \bibfield  {author} {\bibinfo {author} {\bibfnamefont {T.~A.}\ \bibnamefont
  {Wellington}}, \bibinfo {author} {\bibfnamefont {B.~A.}\ \bibnamefont
  {Palles}}, \bibinfo {author} {\bibfnamefont {J.~A.}\ \bibnamefont {Mullens}},
  \bibinfo {author} {\bibfnamefont {J.~T.}\ \bibnamefont {Mihalczo}}, \bibinfo
  {author} {\bibfnamefont {D.~E.}\ \bibnamefont {Archer}}, \bibinfo {author}
  {\bibfnamefont {T.}~\bibnamefont {Thompson}}, \bibinfo {author}
  {\bibfnamefont {C.~L.}\ \bibnamefont {Britton}}, \bibinfo {author}
  {\bibfnamefont {N.~D.~B.}\ \bibnamefont {Ezell}}, \bibinfo {author}
  {\bibfnamefont {M.~N.}\ \bibnamefont {Ericson}}, \bibinfo {author}
  {\bibfnamefont {E.}~\bibnamefont {Farquhar}}, \bibinfo {author}
  {\bibfnamefont {R.}~\bibnamefont {Lind}},\ and\ \bibinfo {author}
  {\bibfnamefont {J.}~\bibnamefont {Carter}},\ }\bibfield  {title} {\enquote
  {\bibinfo {title} {Recent fast neutron imaging measurements with the
  fieldable nuclear materials identification system1},}\ }\href
  {https://doi.org/10.1016/j.phpro.2015.05.054} {\bibfield  {journal} {\bibinfo
   {journal} {Phys. Procedia}\ }\textbf {\bibinfo {volume} {66}},\ \bibinfo
  {pages} {432--438} (\bibinfo {year} {2015})}\BibitemShut {NoStop}%
\bibitem [{\citenamefont {Alexakhin}\ \emph {et~al.}(2015)\citenamefont
  {Alexakhin}, \citenamefont {Bystritsky}, \citenamefont {Zamyatin},
  \citenamefont {Zubarev}, \citenamefont {Krasnoperov}, \citenamefont
  {Rapatsky}, \citenamefont {Rogov}, \citenamefont {Sadovsky}, \citenamefont
  {Salamatin}, \citenamefont {Salmin}, \citenamefont {Sapozhnikov},
  \citenamefont {Slepnev}, \citenamefont {Khabarov}, \citenamefont {Razinkov},
  \citenamefont {Tarasov},\ and\ \citenamefont {Nikitin}}]{ALEAKHIN20159}%
  \BibitemOpen
  \bibfield  {author} {\bibinfo {author} {\bibfnamefont {V.~Y.}\ \bibnamefont
  {Alexakhin}}, \bibinfo {author} {\bibfnamefont {V.~M.}\ \bibnamefont
  {Bystritsky}}, \bibinfo {author} {\bibfnamefont {N.~I.}\ \bibnamefont
  {Zamyatin}}, \bibinfo {author} {\bibfnamefont {E.~V.}\ \bibnamefont
  {Zubarev}}, \bibinfo {author} {\bibfnamefont {A.~V.}\ \bibnamefont
  {Krasnoperov}}, \bibinfo {author} {\bibfnamefont {V.~L.}\ \bibnamefont
  {Rapatsky}}, \bibinfo {author} {\bibfnamefont {Y.~N.}\ \bibnamefont {Rogov}},
  \bibinfo {author} {\bibfnamefont {A.~B.}\ \bibnamefont {Sadovsky}}, \bibinfo
  {author} {\bibfnamefont {A.~V.}\ \bibnamefont {Salamatin}}, \bibinfo {author}
  {\bibfnamefont {R.~A.}\ \bibnamefont {Salmin}}, \bibinfo {author}
  {\bibfnamefont {M.~G.}\ \bibnamefont {Sapozhnikov}}, \bibinfo {author}
  {\bibfnamefont {V.~M.}\ \bibnamefont {Slepnev}}, \bibinfo {author}
  {\bibfnamefont {S.~V.}\ \bibnamefont {Khabarov}}, \bibinfo {author}
  {\bibfnamefont {E.~A.}\ \bibnamefont {Razinkov}}, \bibinfo {author}
  {\bibfnamefont {O.~G.}\ \bibnamefont {Tarasov}},\ and\ \bibinfo {author}
  {\bibfnamefont {G.~M.}\ \bibnamefont {Nikitin}},\ }\bibfield  {title}
  {\enquote {\bibinfo {title} {Detection of diamonds in kimberlite by the
  tagged neutron method},}\ }\href {https://doi.org/10.1016/j.nima.2015.02.049}
  {\bibfield  {journal} {\bibinfo  {journal} {Nucl. Instrum. Methods Phys. Res.
  A}\ }\textbf {\bibinfo {volume} {785}},\ \bibinfo {pages} {9--13} (\bibinfo
  {year} {2015})}\BibitemShut {NoStop}%
\bibitem [{\citenamefont {Litvak}\ \emph {et~al.}(2019)\citenamefont {Litvak},
  \citenamefont {Barmakov}, \citenamefont {Belichenko}, \citenamefont
  {Bestaev}, \citenamefont {Bogolubov}, \citenamefont {Gavrychenkov},
  \citenamefont {Kozyrev}, \citenamefont {Mitrofanov}, \citenamefont {Nosov},
  \citenamefont {Sanin}, \citenamefont {Shvetsov}, \citenamefont {Yurkov},\
  and\ \citenamefont {Zverev}}]{litvak2019}%
  \BibitemOpen
  \bibfield  {author} {\bibinfo {author} {\bibfnamefont {M.}~\bibnamefont
  {Litvak}}, \bibinfo {author} {\bibfnamefont {Y.}~\bibnamefont {Barmakov}},
  \bibinfo {author} {\bibfnamefont {S.}~\bibnamefont {Belichenko}}, \bibinfo
  {author} {\bibfnamefont {R.}~\bibnamefont {Bestaev}}, \bibinfo {author}
  {\bibfnamefont {E.}~\bibnamefont {Bogolubov}}, \bibinfo {author}
  {\bibfnamefont {A.}~\bibnamefont {Gavrychenkov}}, \bibinfo {author}
  {\bibfnamefont {A.}~\bibnamefont {Kozyrev}}, \bibinfo {author} {\bibfnamefont
  {I.}~\bibnamefont {Mitrofanov}}, \bibinfo {author} {\bibfnamefont
  {A.}~\bibnamefont {Nosov}}, \bibinfo {author} {\bibfnamefont
  {A.}~\bibnamefont {Sanin}}, \bibinfo {author} {\bibfnamefont
  {V.}~\bibnamefont {Shvetsov}}, \bibinfo {author} {\bibfnamefont
  {D.}~\bibnamefont {Yurkov}},\ and\ \bibinfo {author} {\bibfnamefont
  {V.}~\bibnamefont {Zverev}},\ }\bibfield  {title} {\enquote {\bibinfo {title}
  {Associated particle imaging instrumentation for future planetary surface
  missions},}\ }\href
  {https://doi.org/https://doi.org/10.1016/j.nima.2018.11.050} {\bibfield
  {journal} {\bibinfo  {journal} {Nucl. Instrum. Methods Phys. Res. A}\
  }\textbf {\bibinfo {volume} {922}},\ \bibinfo {pages} {19 -- 27} (\bibinfo
  {year} {2019})}\BibitemShut {NoStop}%
\bibitem [{\citenamefont {Sanderman}, \citenamefont {Hengl},\ and\
  \citenamefont {Fiske}(2017)}]{Sanderman2017-yz}%
  \BibitemOpen
  \bibfield  {author} {\bibinfo {author} {\bibfnamefont {J.}~\bibnamefont
  {Sanderman}}, \bibinfo {author} {\bibfnamefont {T.}~\bibnamefont {Hengl}},\
  and\ \bibinfo {author} {\bibfnamefont {G.~J.}\ \bibnamefont {Fiske}},\
  }\bibfield  {title} {\enquote {\bibinfo {title} {Soil carbon debt of 12,000
  years of human land use},}\ }\href {https://doi.org/10.1073/pnas.1706103114}
  {\bibfield  {journal} {\bibinfo  {journal} {Proc. Natl. Acad. Sci. U. S. A.}\
  }\textbf {\bibinfo {volume} {114}},\ \bibinfo {pages} {9575--9580} (\bibinfo
  {year} {2017})}\BibitemShut {NoStop}%
\bibitem [{\citenamefont {Huang}\ \emph {et~al.}(2018)\citenamefont {Huang},
  \citenamefont {Minasny}, \citenamefont {McBratney}, \citenamefont
  {Padarian},\ and\ \citenamefont {Triantafilis}}]{Huang2018-md}%
  \BibitemOpen
  \bibfield  {author} {\bibinfo {author} {\bibfnamefont {J.}~\bibnamefont
  {Huang}}, \bibinfo {author} {\bibfnamefont {B.}~\bibnamefont {Minasny}},
  \bibinfo {author} {\bibfnamefont {A.~B.}\ \bibnamefont {McBratney}}, \bibinfo
  {author} {\bibfnamefont {J.}~\bibnamefont {Padarian}},\ and\ \bibinfo
  {author} {\bibfnamefont {J.}~\bibnamefont {Triantafilis}},\ }\bibfield
  {title} {\enquote {\bibinfo {title} {The location- and scale- specific
  correlation between temperature and soil carbon sequestration across the
  globe},}\ }\href {https://doi.org/10.1016/j.scitotenv.2017.09.136} {\bibfield
   {journal} {\bibinfo  {journal} {Sci. Total Environ.}\ }\textbf {\bibinfo
  {volume} {615}},\ \bibinfo {pages} {540--548} (\bibinfo {year}
  {2018})}\BibitemShut {NoStop}%
\bibitem [{\citenamefont {Veres}(2001)}]{LOI}%
  \BibitemOpen
  \bibfield  {author} {\bibinfo {author} {\bibfnamefont {D.-S.}\ \bibnamefont
  {Veres}},\ }\href@noop {} {\enquote {\bibinfo {title} {A comparative study
  between loss on ignition and total carbon analysis on late glacial sediments
  from atteköps mosse, southwestern sweden, and their tentative correlation
  with the grip event stratigraphy},}\ } (\bibinfo {year} {2001}),\ \bibinfo
  {note} {student Paper}\BibitemShut {NoStop}%
\bibitem [{\citenamefont {{ARPA-e ROOTS program}}(2018)}]{ROOTS}%
  \BibitemOpen
  \bibfield  {author} {\bibinfo {author} {\bibnamefont {{ARPA-e ROOTS
  program}}},\ }\href@noop {} {}\bibinfo {howpublished}
  {\url{https://arpa-e.energy.gov/?q=arpa-e-programs/roots}} (\bibinfo {year}
  {2018})\BibitemShut {NoStop}%
\bibitem [{\citenamefont {{Wielopolski}}\ \emph {et~al.}(2000)\citenamefont
  {{Wielopolski}}, \citenamefont {{Orion}}, \citenamefont {{Hendrey}},\ and\
  \citenamefont {{Roger}}}]{BNLsoil}%
  \BibitemOpen
  \bibfield  {author} {\bibinfo {author} {\bibfnamefont {L.}~\bibnamefont
  {{Wielopolski}}}, \bibinfo {author} {\bibfnamefont {I.}~\bibnamefont
  {{Orion}}}, \bibinfo {author} {\bibfnamefont {G.}~\bibnamefont {{Hendrey}}},\
  and\ \bibinfo {author} {\bibfnamefont {H.}~\bibnamefont {{Roger}}},\
  }\bibfield  {title} {\enquote {\bibinfo {title} {Soil carbon measurements
  using inelastic neutron scattering},}\ }\href
  {https://doi.org/10.1109/23.856717} {\bibfield  {journal} {\bibinfo
  {journal} {IEEE Transactions on Nuclear Science}\ }\textbf {\bibinfo {volume}
  {47}},\ \bibinfo {pages} {914--917} (\bibinfo {year} {2000})}\BibitemShut
  {NoStop}%
\bibitem [{\citenamefont {Wielopolski}\ \emph {et~al.}(2008)\citenamefont
  {Wielopolski}, \citenamefont {Hendrey}, \citenamefont {Johnsen},
  \citenamefont {Mitra}, \citenamefont {Prior}, \citenamefont {Rogers},\ and\
  \citenamefont {Torbert}}]{Wielopolski2008-ff}%
  \BibitemOpen
  \bibfield  {author} {\bibinfo {author} {\bibfnamefont {L.}~\bibnamefont
  {Wielopolski}}, \bibinfo {author} {\bibfnamefont {G.}~\bibnamefont
  {Hendrey}}, \bibinfo {author} {\bibfnamefont {K.~H.}\ \bibnamefont
  {Johnsen}}, \bibinfo {author} {\bibfnamefont {S.}~\bibnamefont {Mitra}},
  \bibinfo {author} {\bibfnamefont {S.~A.}\ \bibnamefont {Prior}}, \bibinfo
  {author} {\bibfnamefont {H.~H.}\ \bibnamefont {Rogers}},\ and\ \bibinfo
  {author} {\bibfnamefont {H.~A.}\ \bibnamefont {Torbert}},\ }\bibfield
  {title} {\enquote {\bibinfo {title} {Nondestructive system for analyzing
  carbon in the soil},}\ }\href {https://doi.org/10.2136/sssaj2007.0177}
  {\bibfield  {journal} {\bibinfo  {journal} {Soil Sci. Soc. Am. J.}\ }\textbf
  {\bibinfo {volume} {72}},\ \bibinfo {pages} {1269--1277} (\bibinfo {year}
  {2008})}\BibitemShut {NoStop}%
\bibitem [{\citenamefont {Kavetskiy}\ \emph
  {et~al.}(2019{\natexlab{a}})\citenamefont {Kavetskiy}, \citenamefont
  {Yakubova}, \citenamefont {Prior},\ and\ \citenamefont
  {Torbert}}]{Kavetskiy2019-rg}%
  \BibitemOpen
  \bibfield  {author} {\bibinfo {author} {\bibfnamefont {A.}~\bibnamefont
  {Kavetskiy}}, \bibinfo {author} {\bibfnamefont {G.}~\bibnamefont {Yakubova}},
  \bibinfo {author} {\bibfnamefont {S.~A.}\ \bibnamefont {Prior}},\ and\
  \bibinfo {author} {\bibfnamefont {H.~A.}\ \bibnamefont {Torbert}},\
  }\bibfield  {title} {\enquote {\bibinfo {title} {Application of associated
  particle neutron techniques for soil carbon analysis},}\ }\href
  {https://doi.org/10.1063/1.5127698} {\bibfield  {journal} {\bibinfo
  {journal} {AIP Conf. Proc.}\ }\textbf {\bibinfo {volume} {2160}},\ \bibinfo
  {pages} {050006} (\bibinfo {year} {2019}{\natexlab{a}})}\BibitemShut
  {NoStop}%
\bibitem [{\citenamefont {Kavetskiy}\ \emph
  {et~al.}(2019{\natexlab{b}})\citenamefont {Kavetskiy}, \citenamefont
  {Yakubova}, \citenamefont {Prior},\ and\ \citenamefont
  {Torbert~(USDA)}}]{Kavetskiy2019-lt}%
  \BibitemOpen
  \bibfield  {author} {\bibinfo {author} {\bibfnamefont {A.}~\bibnamefont
  {Kavetskiy}}, \bibinfo {author} {\bibfnamefont {G.}~\bibnamefont {Yakubova}},
  \bibinfo {author} {\bibfnamefont {S.~A.}\ \bibnamefont {Prior}},\ and\
  \bibinfo {author} {\bibfnamefont {H.~A.}\ \bibnamefont {Torbert~(USDA)}},\
  }\bibfield  {title} {\enquote {\bibinfo {title} {Application of
  {Neutron-Gamma} technologies in agriculture},}\ }\href
  {https://doi.org/10.13182/T31237} {\bibfield  {journal} {\bibinfo  {journal}
  {TRANS}\ }\textbf {\bibinfo {volume} {121}},\ \bibinfo {pages} {539--541}
  (\bibinfo {year} {2019}{\natexlab{b}})}\BibitemShut {NoStop}%
\bibitem [{\citenamefont {Yakubova}\ \emph {et~al.}(2016)\citenamefont
  {Yakubova}, \citenamefont {Kavetskiy}, \citenamefont {Prior},\ and\
  \citenamefont {Torbert}}]{Yakubova2016-tk}%
  \BibitemOpen
  \bibfield  {author} {\bibinfo {author} {\bibfnamefont {G.}~\bibnamefont
  {Yakubova}}, \bibinfo {author} {\bibfnamefont {A.}~\bibnamefont {Kavetskiy}},
  \bibinfo {author} {\bibfnamefont {S.~A.}\ \bibnamefont {Prior}},\ and\
  \bibinfo {author} {\bibfnamefont {H.~A.}\ \bibnamefont {Torbert}},\
  }\bibfield  {title} {\enquote {\bibinfo {title} {Benchmarking the inelastic
  neutron scattering soil carbon method},}\ }\href
  {https://doi.org/10.2136/vzj2015.04.0056} {\bibfield  {journal} {\bibinfo
  {journal} {Vadose Zone J.}\ }\textbf {\bibinfo {volume} {15}} (\bibinfo
  {year} {2016}),\ 10.2136/vzj2015.04.0056}\BibitemShut {NoStop}%
\bibitem [{\citenamefont {{Adelphi Technology Inc.}}(2018)}]{Adelphi}%
  \BibitemOpen
  \bibfield  {author} {\bibinfo {author} {\bibnamefont {{Adelphi Technology
  Inc.}}},\ }\href@noop {} {}\bibinfo {howpublished}
  {\url{http://adelphitech.com/}} (\bibinfo {year} {2018})\BibitemShut
  {NoStop}%
\bibitem [{\citenamefont {{CRYTUR, spol. s r.o.}}(2018)}]{Crytur}%
  \BibitemOpen
  \bibfield  {author} {\bibinfo {author} {\bibnamefont {{CRYTUR, spol. s
  r.o.}}},\ }\href@noop {} {}\bibinfo {howpublished} {\url{https://crytur.cz}}
  (\bibinfo {year} {2018})\BibitemShut {NoStop}%
\bibitem [{\citenamefont {{Hamamatsu Photonics K.K.}}(2018)}]{Hamamatsu}%
  \BibitemOpen
  \bibfield  {author} {\bibinfo {author} {\bibnamefont {{Hamamatsu Photonics
  K.K.}}},\ }\href@noop {} {}\bibinfo {howpublished}
  {\url{https://www.hamamatsu.com/us/en/product/type/H13700/index.html}}
  (\bibinfo {year} {2018})\BibitemShut {NoStop}%
\bibitem [{\citenamefont {{XIA LLC}}(2018)}]{XIA}%
  \BibitemOpen
  \bibfield  {author} {\bibinfo {author} {\bibnamefont {{XIA LLC}}},\
  }\href@noop {} {}\bibinfo {howpublished} {\url{https://xia.com}} (\bibinfo
  {year} {2018})\BibitemShut {NoStop}%
\bibitem [{\citenamefont {Unzueta}\ \emph {et~al.}(2019)\citenamefont
  {Unzueta}, \citenamefont {Mixter}, \citenamefont {Croft}, \citenamefont
  {Joseph}, \citenamefont {Ludewigt},\ and\ \citenamefont
  {Persaud}}]{Mauricio1}%
  \BibitemOpen
  \bibfield  {author} {\bibinfo {author} {\bibfnamefont {M.~A.}\ \bibnamefont
  {Unzueta}}, \bibinfo {author} {\bibfnamefont {W.}~\bibnamefont {Mixter}},
  \bibinfo {author} {\bibfnamefont {Z.}~\bibnamefont {Croft}}, \bibinfo
  {author} {\bibfnamefont {J.}~\bibnamefont {Joseph}}, \bibinfo {author}
  {\bibfnamefont {B.}~\bibnamefont {Ludewigt}},\ and\ \bibinfo {author}
  {\bibfnamefont {A.}~\bibnamefont {Persaud}},\ }\bibfield  {title} {\enquote
  {\bibinfo {title} {Position sensitive alpha detector for an associated
  particle imaging system},}\ }\href {https://doi.org/10.1063/1.5127697}
  {\bibfield  {journal} {\bibinfo  {journal} {AIP Conference Proceedings}\
  }\textbf {\bibinfo {volume} {2160}},\ \bibinfo {pages} {050005} (\bibinfo
  {year} {2019})},\ \Eprint {https://arxiv.org/abs/1811.08591}
  {arXiv:1811.08591} \BibitemShut {NoStop}%
\bibitem [{\citenamefont {Moszy{\'n}ski}\ \emph {et~al.}(1998)\citenamefont
  {Moszy{\'n}ski}, \citenamefont {Kapusta}, \citenamefont {Wolski},
  \citenamefont {Klamra},\ and\ \citenamefont {Cederwall}}]{Moszynski1998-hh}%
  \BibitemOpen
  \bibfield  {author} {\bibinfo {author} {\bibfnamefont {M.}~\bibnamefont
  {Moszy{\'n}ski}}, \bibinfo {author} {\bibfnamefont {M.}~\bibnamefont
  {Kapusta}}, \bibinfo {author} {\bibfnamefont {D.}~\bibnamefont {Wolski}},
  \bibinfo {author} {\bibfnamefont {W.}~\bibnamefont {Klamra}},\ and\ \bibinfo
  {author} {\bibfnamefont {B.}~\bibnamefont {Cederwall}},\ }\bibfield  {title}
  {\enquote {\bibinfo {title} {Properties of the {YAP} : Ce scintillator},}\
  }\href {https://doi.org/10.1016/S0168-9002(97)01115-7} {\bibfield  {journal}
  {\bibinfo  {journal} {Nucl. Instrum. Methods Phys. Res. A}\ }\textbf
  {\bibinfo {volume} {404}},\ \bibinfo {pages} {157--165} (\bibinfo {year}
  {1998})}\BibitemShut {NoStop}%
\bibitem [{\citenamefont {{COMSOL Multiphysics v. 5.2}}(2018)}]{Comsol}%
  \BibitemOpen
  \bibfield  {author} {\bibinfo {author} {\bibnamefont {{COMSOL Multiphysics v.
  5.2}}},\ }\href@noop {} {}\bibinfo {howpublished}
  {\url{https://www.comsol.com}} (\bibinfo {year} {2018})\BibitemShut {NoStop}%
\bibitem [{\citenamefont {{Refractive Index
  Database}}(2019)}]{refractiveIndex}%
  \BibitemOpen
  \bibfield  {author} {\bibinfo {author} {\bibnamefont {{Refractive Index
  Database}}},\ }\href@noop {} {}\bibinfo {howpublished}
  {https://refractiveindex.info} (\bibinfo {year} {2019})\BibitemShut {NoStop}%
\bibitem [{\citenamefont {{LTspice from Analog Devices}}(2018)}]{LTSpice}%
  \BibitemOpen
  \bibfield  {author} {\bibinfo {author} {\bibnamefont {{LTspice from Analog
  Devices}}},\ }\href@noop {} {}\bibinfo {howpublished}
  {\url{https://www.analog.com/LTspice}} (\bibinfo {year} {2018})\BibitemShut
  {NoStop}%
\bibitem [{\citenamefont {Unzueta}\ \emph {et~al.}(2020)\citenamefont
  {Unzueta}, \citenamefont {Tak}, \citenamefont {Ludewigt},\ and\ \citenamefont
  {Persaud}}]{data}%
  \BibitemOpen
  \bibfield  {author} {\bibinfo {author} {\bibfnamefont {M.~A.}\ \bibnamefont
  {Unzueta}}, \bibinfo {author} {\bibfnamefont {T.}~\bibnamefont {Tak}},
  \bibinfo {author} {\bibfnamefont {B.}~\bibnamefont {Ludewigt}},\ and\
  \bibinfo {author} {\bibfnamefont {A.}~\bibnamefont {Persaud}},\ }\bibfield
  {title} {\enquote {\bibinfo {title} {Data, analysis scripts, and simulations
  files for 'an all-digital associated particle imaging system for the 3d
  determinationof isotopic distributions'},}\ }\href
  {https://doi.org/10.5281/zenodo.4737896} {10.5281/zenodo.4737896} (\bibinfo
  {year} {2020}),\ \bibinfo {note} {{Zenodo}}\BibitemShut {NoStop}%
\bibitem [{\citenamefont {Zhang}\ \emph {et~al.}(2012)\citenamefont {Zhang},
  \citenamefont {Hayward}, \citenamefont {Cates}, \citenamefont {Hausladen},
  \citenamefont {Laubach}, \citenamefont {Sparger},\ and\ \citenamefont
  {Donnald}}]{Zhang2012}%
  \BibitemOpen
  \bibfield  {author} {\bibinfo {author} {\bibfnamefont {X.}~\bibnamefont
  {Zhang}}, \bibinfo {author} {\bibfnamefont {J.~P.}\ \bibnamefont {Hayward}},
  \bibinfo {author} {\bibfnamefont {J.~W.}\ \bibnamefont {Cates}}, \bibinfo
  {author} {\bibfnamefont {P.~A.}\ \bibnamefont {Hausladen}}, \bibinfo {author}
  {\bibfnamefont {M.~A.}\ \bibnamefont {Laubach}}, \bibinfo {author}
  {\bibfnamefont {J.~E.}\ \bibnamefont {Sparger}},\ and\ \bibinfo {author}
  {\bibfnamefont {S.~B.}\ \bibnamefont {Donnald}},\ }\bibfield  {title}
  {\enquote {\bibinfo {title} {Benchmarking the geant4 full system simulation
  of an associated alpha-particle detector for use in a d–t neutron
  generator},}\ }\href
  {https://doi.org/https://doi.org/10.1016/j.apradiso.2012.04.026} {\bibfield
  {journal} {\bibinfo  {journal} {Applied Radiation and Isotopes}\ }\textbf
  {\bibinfo {volume} {70}},\ \bibinfo {pages} {1485 -- 1493} (\bibinfo {year}
  {2012})}\BibitemShut {NoStop}%
\bibitem [{\citenamefont {Ayllon~Unzueta}(2020)}]{thesisMauricio}%
  \BibitemOpen
  \bibfield  {author} {\bibinfo {author} {\bibfnamefont {M.}~\bibnamefont
  {Ayllon~Unzueta}},\ }\emph {\bibinfo {title} {An Associated Particle Imaging
  System for the Determination of 3D Isotopic Distributions}},\ \href@noop {}
  {Ph.D. thesis},\ \bibinfo  {school} {UC Berkeley} (\bibinfo {year}
  {2020})\BibitemShut {NoStop}%
\bibitem [{\citenamefont {Goorley}\ \emph {et~al.}(2012)\citenamefont
  {Goorley}, \citenamefont {James}, \citenamefont {Booth}, \citenamefont
  {Brown}, \citenamefont {Bull}, \citenamefont {Cox}, \citenamefont {Durkee},
  \citenamefont {Elson}, \citenamefont {Fensin}, \citenamefont {Forster} \emph
  {et~al.}}]{MCNP}%
  \BibitemOpen
  \bibfield  {author} {\bibinfo {author} {\bibfnamefont {T.}~\bibnamefont
  {Goorley}}, \bibinfo {author} {\bibfnamefont {M.}~\bibnamefont {James}},
  \bibinfo {author} {\bibfnamefont {T.}~\bibnamefont {Booth}}, \bibinfo
  {author} {\bibfnamefont {F.}~\bibnamefont {Brown}}, \bibinfo {author}
  {\bibfnamefont {J.}~\bibnamefont {Bull}}, \bibinfo {author} {\bibfnamefont
  {L.}~\bibnamefont {Cox}}, \bibinfo {author} {\bibfnamefont {J.}~\bibnamefont
  {Durkee}}, \bibinfo {author} {\bibfnamefont {J.}~\bibnamefont {Elson}},
  \bibinfo {author} {\bibfnamefont {M.}~\bibnamefont {Fensin}}, \bibinfo
  {author} {\bibfnamefont {R.}~\bibnamefont {Forster}}, \emph {et~al.},\
  }\bibfield  {title} {\enquote {\bibinfo {title} {Initial mcnp6 release
  overview},}\ }\href {https://doi.org/10.13182/NT11-135} {\bibfield  {journal}
  {\bibinfo  {journal} {Nuclear Technology}\ }\textbf {\bibinfo {volume}
  {180}},\ \bibinfo {pages} {298--315} (\bibinfo {year} {2012})}\BibitemShut
  {NoStop}%
\bibitem [{\citenamefont {et. al.}(2018)}]{ENDF}%
  \BibitemOpen
  \bibfield  {author} {\bibinfo {author} {\bibfnamefont {D.~B.}\ \bibnamefont
  {et. al.}},\ }\bibfield  {title} {\enquote {\bibinfo {title} {Endf/b-viii.0:
  The 8th major release of the nuclear reaction data library with cielo-project
  cross sections, new standards and thermal scattering data},}\ }\href
  {https://doi.org/https://doi.org/10.1016/j.nds.2018.02.001} {\bibfield
  {journal} {\bibinfo  {journal} {Nuclear Data Sheets}\ }\textbf {\bibinfo
  {volume} {148}},\ \bibinfo {pages} {1 -- 142} (\bibinfo {year} {2018})},\
  \bibinfo {note} {special Issue on Nuclear Reaction Data}\BibitemShut
  {NoStop}%
\bibitem [{\citenamefont {{Mauborgne, Marie-Laure}}\ \emph
  {et~al.}(2017)\citenamefont {{Mauborgne, Marie-Laure}}, \citenamefont
  {{Allioli, Fran\c{c}oise}}, \citenamefont {{Manclossi, Mauro}}, \citenamefont
  {{Nicoletti, Luisa}}, \citenamefont {{Stoller, Chris}},\ and\ \citenamefont
  {{Evans, Mike}}}]{Mauborgne}%
  \BibitemOpen
  \bibfield  {author} {\bibinfo {author} {\bibnamefont {{Mauborgne,
  Marie-Laure}}}, \bibinfo {author} {\bibnamefont {{Allioli, Fran\c{c}oise}}},
  \bibinfo {author} {\bibnamefont {{Manclossi, Mauro}}}, \bibinfo {author}
  {\bibnamefont {{Nicoletti, Luisa}}}, \bibinfo {author} {\bibnamefont
  {{Stoller, Chris}}},\ and\ \bibinfo {author} {\bibnamefont {{Evans, Mike}}},\
  }\bibfield  {title} {\enquote {\bibinfo {title} {Designing tools for oil
  exploration using nuclear modeling},}\ }\href
  {https://doi.org/10.1051/epjconf/201714609036} {\bibfield  {journal}
  {\bibinfo  {journal} {EPJ Web Conf.}\ }\textbf {\bibinfo {volume} {146}},\
  \bibinfo {pages} {09036} (\bibinfo {year} {2017})}\BibitemShut {NoStop}%
\bibitem [{\citenamefont {Cates}, \citenamefont {Hayward},\ and\ \citenamefont
  {Zhang}(2013)}]{Cates2013}%
  \BibitemOpen
  \bibfield  {author} {\bibinfo {author} {\bibfnamefont {J.~W.}\ \bibnamefont
  {Cates}}, \bibinfo {author} {\bibfnamefont {J.~P.}\ \bibnamefont {Hayward}},\
  and\ \bibinfo {author} {\bibfnamefont {X.}~\bibnamefont {Zhang}},\ }\bibfield
   {title} {\enquote {\bibinfo {title} {Achievable position resolution of an
  alpha detector with continuous spatial response for use in associated
  particle imaging},}\ }in\ \href {https://doi.org/10.1109/NSSMIC.2013.6829782}
  {\emph {\bibinfo {booktitle} {2013 IEEE Nuclear Science Symposium and Medical
  Imaging Conference (2013 NSS/MIC)}}}\ (\bibinfo {year} {2013})\ pp.\ \bibinfo
  {pages} {1--3}\BibitemShut {NoStop}%
\bibitem [{\citenamefont {Warburton}\ and\ \citenamefont
  {Hennig}(2017)}]{Warburton2017-qe}%
  \BibitemOpen
  \bibfield  {author} {\bibinfo {author} {\bibfnamefont {W.~K.}\ \bibnamefont
  {Warburton}}\ and\ \bibinfo {author} {\bibfnamefont {W.}~\bibnamefont
  {Hennig}},\ }\bibfield  {title} {\enquote {\bibinfo {title} {New algorithms
  for improved digital pulse arrival timing with {Sub-GSps} {ADCs}},}\ }\href
  {https://doi.org/10.1109/TNS.2017.2766074} {\bibfield  {journal} {\bibinfo
  {journal} {IEEE Trans. Nucl. Sci.}\ }\textbf {\bibinfo {volume} {64}},\
  \bibinfo {pages} {2938--2950} (\bibinfo {year} {2017})}\BibitemShut {NoStop}%
\end{thebibliography}%
\end{document}